\documentclass[journal]{IEEEtran}
\usepackage{amsmath,amssymb,amsfonts}
\usepackage{algorithmic}
\usepackage{graphicx}
\usepackage{textcomp}
\usepackage{xcolor}
\usepackage{pifont}
\usepackage{color,soul}
\usepackage{dblfloatfix}
\usepackage{tabularray}
\usepackage{xcolor}
\usepackage{longtable}
\usepackage{lettrine}
\usepackage{fontawesome5}
\usepackage{caption}
\usepackage{multirow}
\usepackage{caption}
\usepackage{subcaption}

\usepackage{xcolor}
\usepackage{graphicx}
\usepackage{colortbl}
\usepackage{array}
\usepackage{amsmath,amssymb}
\usepackage{tikz}
\usepackage{graphics}
\usepackage{bbding}
\newcommand{\cmark}{\text{\ding{51}}}
\newcommand{\xmark}{\text{\ding{55}}}

\def\BibTeX{{\rm B\kern-.05em{\sc i\kern-.025em b}\kern-.08em
 T\kern-.1667em\lower.7ex\hbox{E}\kern-.125emX}}

\begin{document}

\title{Privacy in Artificial Intelligence Extended Reality (AI-XR) Enabled Metaverses: Risks and Solutions}
\title{Privacy Protection in the Metaverse:\\ Addressing AI and Extended Reality Threats}
\title{Privacy Preservation in the Metaverse:\\ Countering The Privacy Threats Posed by Artificial Intelligence (AI) and Extended Reality (XR)}
\title{Privacy Preservation in the AI-XR Metaverse:\\ Countering The Privacy Threats Posed by Artificial Intelligence (AI) and Extended Reality (XR)}
\title{Privacy Preservation in Artificial Intelligence and Extended Reality (AI-XR) Metaverses: A Survey}

\author{Mahdi Alkaeed$^{1}$, Adnan Qayyum$^{2}$, Junaid Qadir$^{1}$ \\
$^1$Department of Computer Science and Engineering, College of Engineering, Qatar University, Doha, Qatar \\
$^2$Information Technology University (ITU), Lahore, Pakistan\\
}


\maketitle
\begin{abstract}
The metaverse is a nascent concept that envisions a virtual universe, a collaborative space where individuals can interact, create, and participate in a wide range of activities. Privacy in the metaverse is a critical concern as the concept evolves and immersive virtual experiences become more prevalent.
The metaverse privacy problem refers to the challenges and concerns surrounding the privacy of personal information and data within  Virtual Reality (VR) environments as the concept of a shared VR space becomes more accessible. Metaverse will harness advancements from various technologies such as Artificial Intelligence (AI), Extended Reality (XR), Mixed Reality (MR), and 5G/6G-based communication to provide personalized and immersive services to its users. Moreover, to enable more personalized experiences, the metaverse relies on the collection of fine-grained user data that leads to various privacy issues.
Therefore, before the potential of the metaverse can be fully realized, privacy concerns related to personal information and data within VR environments must be addressed. This includes safeguarding users' control over their data, ensuring the security of their personal information, and protecting in-world actions and interactions from unauthorized sharing. In this paper, we explore various privacy challenges that future metaverses are expected to face, given their reliance on AI for tracking users, creating XR and MR experiences, and facilitating interactions. Moreover, we thoroughly analyze technical solutions such as differential privacy, Homomorphic Encryption (HE), and Federated Learning (FL) and discuss related sociotechnical issues regarding privacy.
\end{abstract}

\begin{IEEEkeywords}
Machine Learning, Metaverse, Artificial Intelligence, Virtual Reality, Extended Reality, Mixed Reality, Homomorphic Encryption, and Federated Learning.
\end{IEEEkeywords}

\section{Introduction}
\lettrine[lines=2]{T}he metaverse refers to a virtual world that combines physical reality with digital technology leveraging Artificial Intelligence (AI) and Extended Reality (XR), creating a network of interconnected and immersive environments. It blends Virtual Reality (VR), Augmented Reality (AR), permitting users to engage in multi-sensory interactions with virtual objects, people, and environments \cite{mystakidis2022Metaverse}. AR is a technology that superimposes digital information, such as images, sounds, and other data, onto the real-world environment in real-time \cite{encyclopedia2010031}. AR enhances the user's perception of the physical world by overlaying computer-generated sensory input such as graphics, audio, and haptic feedback. Mixed Reality (MR) is a form of immersive technology that blends virtual and physical worlds to create a new reality that allows users to interact with digital objects and environments as if they were real. VR completely immerses users in a simulated environment, whereas MR allows users to interact with both the physical world and virtual objects simultaneously \cite{FLAVIAN2019547}. In the metaverse, users can engage in real-time communication and dynamic interactions, such as socializing with friends, playing multiplayer games, and exploring open virtual worlds. 

However, as the concept of a metaverse becomes more prevalent in our society, privacy concerns arising from the collection and utilization of fine-grained personal information by metaverse creators and platforms---e.g., users' location, browsing history, personal preferences \cite{zhao2022metaverse}---assume grave importance \cite{wang2022survey}. The metaverse may collect a range of essential information from its users, including user profile information, interaction data, biometric data, payment information, and device information \cite{10026513}. Users may need to provide payment information to purchase virtual items and assets or access premium features, and device information such as IP addresses, device IDs, and operating system details may also be collected \cite{wang2022survey}. Biometric data such as facial recognition or voice prints may be collected to verify a user's identity or enable voice chat. Such data can be used for targeted advertising, data mining, and other purposes. Additionally, metaverse data could be accessed by cybercriminals maliciously through hacking posing potential harm to individuals through privacy violation.

\begin{table*}[!h]
\centering
\caption{ Comparison of our paper with existing similar survey and review articles. \\ }
\label{tab:contributionsTable}
\scalebox{0.8}{
\begin{tabular} {|p{0.5cm}|p{2.2cm}|p{3cm}|p{7.5cm}|p{1cm}|p{1cm}|p{1cm}|p{1cm}|p{0.9cm}|}
\hline
\textbf{Year}  & \textbf{Reference}& \textbf{Focused Area } & \textbf{Contribution(s)} & \textbf{AI\newline Privacy (AIP)} &
\textbf{Metaverse\newline Privacy (MP)} 
&
\textbf{AIP\newline Solutions} 
&
\textbf{MP\newline Solutions} 
&
\textbf{Open Issues}\\ \hline
2018
& 
Falchuk et al.~\cite{falchuk2018social}
&
Social metaverse games
&
Discussed general investigating privacy concerns and solutions associated with digital footprints in social metaverse games.
&
\xmark 
&
\cmark 
& 
\xmark
&
\cmark 
&
\cmark
\\
\hline
2021
&
Ning et al.~\cite{ning2021survey} 
& 
Metaverse industry.
&
Analyzed the progress and advancements in the metaverse industry, covering national policies, technological infrastructure, VR, and social metaverse platforms.
&
\xmark 
&
\cmark 
& 
\xmark
&
\xmark
&
\cmark
\\
\hline
2022
&
Yang et al.~\cite{yang2022fusing} 
&
AI-extended metaverse
&
Analyzed leveraging AI and privacy-preserving blockchain technologies for building the future of the metaverse.
&
\xmark
&
\cmark 
& 
\xmark
&
\cmark 
&
\cmark
\\
\hline
2022
&
Xu at al. \cite{xu2022full}
&
Metaverse communication networks
&
Explored how edge computing supports the metaverse through communication networks and computational capabilities.
&
\xmark  
&
\cmark
&
\xmark
&
\cmark
&
\cmark
\\
\hline
2022
&
Xu et al.~\cite{xu2022full}, 
&
Edge-enabled metaverse
&
Comprehensive analysis of the edge-enabled metaverse, with a focus on various aspects such as communication, networking, computation, and blockchain technology.
&
\xmark
&
\cmark 
& 
\xmark
&
\cmark
&
\cmark
\\
\hline
2022
&
Zhao et al.~\cite{zhao2022metaverse}
&
Security and privacy challenges in the metaverse.
&
Several potential solutions addressing the security and privacy concerns in the metaverse
&
\xmark  
&
\cmark 
&
\xmark
&
\cmark
&
\cmark
\\

\hline
2023
&
Wang et al.~\cite{wang2022survey}
&
Metaverse systems
&
Examined the security and privacy threats that metaverse systems pose and addresses the significant challenges that arise as a result.
&
\xmark  
&
\cmark 
& 
\xmark
&
\cmark
&
\cmark
\\
\hline
2023
&
Huynh et al.~\cite{huynh2023artificial}
&
AI-extended metaverse
&
How AI techniques can contribute to the establishment and advancement of the metaverse.
&
\xmark 
&
\xmark
& 
\xmark
&
\xmark
&
\cmark
\\
\hline
2023
&
This survey.
&
Impact of data collection in AI-XR enabled metaverse
Anonymization techniques.
&
This survey includes an in-depth analysis of the metaverse architecture and associated privacy risks. The survey also explores major challenges and innovative solutions, as well as potential research areas to ensure a secure and protected metaverse.
&
\cmark   
&
\cmark 
& 
\cmark
&
\cmark
&
\cmark
\\
\hline
\end{tabular}}
\label{tab:table_label}
\end{table*}

To address these concerns, AI-XR platforms and creators should adopt strict privacy policies and implement measures for robust security, including encryption, and two-factor authentication to protect user data from unauthorized access. To protect user privacy in the metaverse, users should be cautious about sharing personal information in the metaverse. Interaction data collected by the metaverse could include information on the communities users join, virtual items they interact with, and purchases they make. Additionally, privacy measures must be regularly reviewed and updated to keep pace with new developments in this constantly evolving technology \cite{park2022Metaverse}.

ML can also play its role in enhancing privacy in AI-XR metaverses in diverse ways (e.g., in developing advanced encryption algorithms, anomaly detection for detecting unusual behavior, and differential privacy techniques that provide strong privacy guarantees by adding noise to the data) to detect malicious activities \cite{wang2022wireless,peterson2019machine}. ML can be utilized to identify and flag suspicious activity \cite{TIANKAI} and detect malicious actions such as spam or phishing attempts \cite{huynh2023artificial}. ML can assist in automating data anonymization and de-identification processes, enabling organizations to protect sensitive information while still utilizing it for analytic and research purposes. ML can also monitor and analyze vast quantities of data in real-time to quickly detect suspicious activity and ensure a secure environment for all participants in AI-XR \cite{zhang2019real,jiao2019tourism}. 

Overall, the paper aims to contribute to the understanding of metaverse privacy challenges and outlines potential solutions, providing insights for researchers, policymakers, and practitioners involved in the development and utilization of metaverse technologies. The paper explores concerns surrounding the privacy of personal information and data within VR environments and discusses the challenges that the metaverse is expected to face in terms of privacy. 

\textit{Contributions of this Paper:} 
\begin{enumerate}
\item We comprehensively discuss various privacy challenges that are anticipated to arise in the future metaverse.
\item We provide a detailed discussion and taxonomy of potential solutions that can be leveraged to mitigate privacy issues associated with AI-XR-enabled metaverses.
\item We explore the intersection of privacy and AI-XR and analyze how the interconnection of these technologies can pose new privacy challenges.
\item  We analyzed the impact of data collection in AI-XR-enabled metaverse and highlight how it can be used for user profiling and tracking--leading to several privacy risks. 
\item  Finally, we elaborate upon various recommendations for different stakeholders that include individuals, organizations, and policymakers to enable them in understanding the enormity of privacy issues and the need to address them using appropriate privacy-preserving techniques.
\end{enumerate} 

Table \ref{tab:contributionsTable} provides a detailed comparison of our paper with existing survey and review articles having similar focus. From the table, it is evident that our work has made significant contributions in this area, by building on and synthesizing the findings of previous studies. Table \ref{tab:notation} presents a summary of the notation used in this paper.

\begin{table}[!h]
\centering
\caption{Table of Notation.}
\label{tab:notation}
\begin{tabular}{|l p{6cm}|}
\hline
\textbf{Notation}  & \textbf{Description} \\ \hline

$C_i$
& 
The encrypted data
\\ 
$K$
& 
The number of edge devices (ED).  
\\ 
 $Enc$,$Dec$
& 
 Encryption, decryption functions.
\\ 
 $M$
& 
A randomized algorithm for DP
\\ 
 $\theta$
& 
 AI-XR metaverse model parameters.
\\ 
 $N$
& 
The number of examples in the labeled dataset.
\\ 
 $L()$
& 
The loss function
\\ 
$t$
& 
The communication round.
\\ 
$\theta_{t+1}$
& 
The updated global model after $t$
\\
$\theta_(i)$
& 
The local model at device $i$
\\
$k$
& 
The number of participants 
\\
$n_k$
& 
The samples of participants $k$
\\
$\theta_{t+1}^k$
&
The local model parameter of $k$
\\ 
$W_i$ 
& 
Local uploading parameters for $i$th ED.
\\ 
 $w$ 
& 
The vector of model parameters.
\\ 
 $X_i$ 
& 
Entries of the database.
\\
 $Y_i$
& 
The true label.
\\
\hline
\end{tabular}

\label{tab:table_label}
\end{table}

\begin{figure*} [!t]
\centering
\includegraphics[width=.9\linewidth]{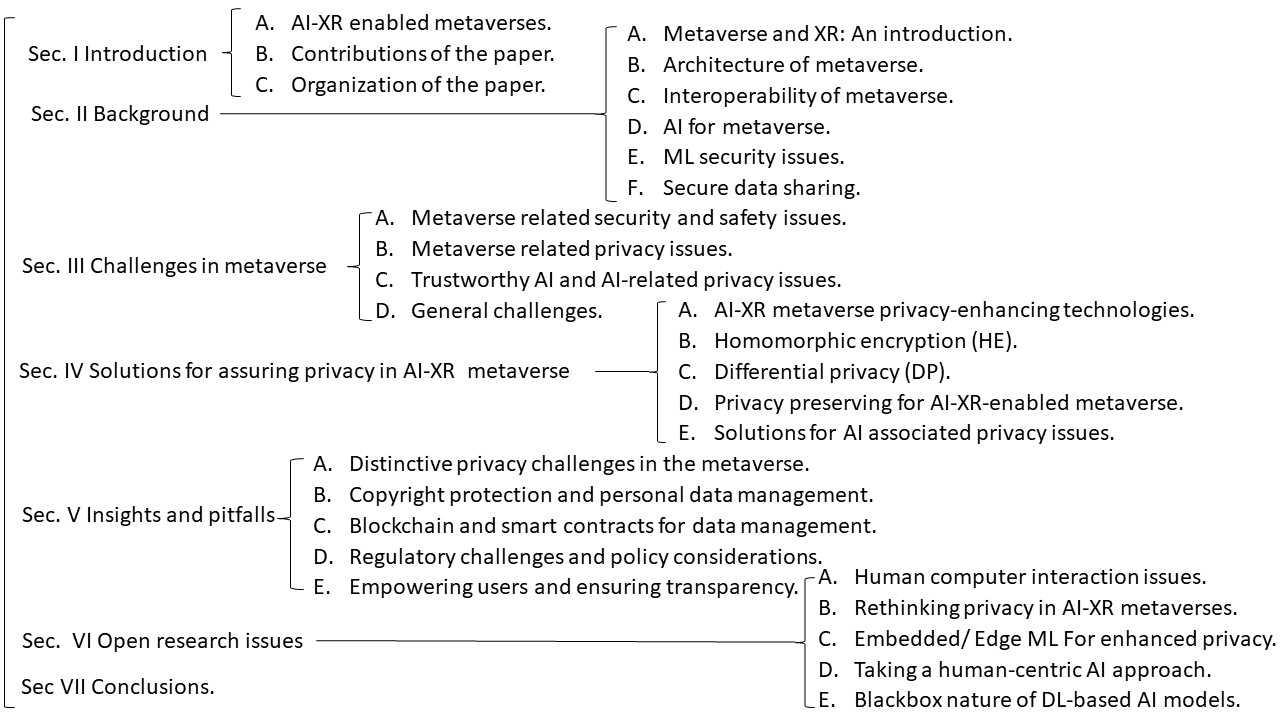}
\caption{Organization of the paper.}
\label{Fig:OrganizationofthePaper}
\end{figure*}

\textit{Organization of this Paper (Figure \ref{Fig:OrganizationofthePaper}):} The rest of this paper is organized as follows: Section \ref{sec:background} provides a comprehensive background on AI-XR-enabled metaverse privacy. Section \ref{sec:Challenges} examines previous studies and works that have investigated privacy concerns related to the metaverse. Section \ref{sec:Solutions} presents solutions for assuring privacy in the metaverse, including a review of privacy policies of metaverse platforms and related work summarized in tables. These tables offer a comprehensive summary of the primary privacy issues linked to the metaverse and ML. Lastly, the paper concludes with Section\ref{sec:conclusions}.

\section{Background}
\label{sec:background}

\subsection{Metaverse and XR: An Introduction}
The metaverse is an emerging concept that describes a fully immersive digital realm in which individuals can engage with both digital objects and each other within a virtual space. The metaverse leverages the concepts of avatar, XR, VR, MR, AR to provide immersive digital experiences, however, they have some key differences.

\subsubsection{Metaverse}
It is often described as a kind of VR internet, where users can navigate a persistent, shared, and often gamified universe. The metaverse is still largely a hypothetical concept, but it is gaining attention as technology progresses and virtual experiences become more advanced.

\subsubsection{Extended Reality (XR)}
The metaverse, closely associated with XR, is a virtual universe or shared space that users can access through various devices and interact with in real-time. XR is an umbrella term for immersive digital experiences that blend physical reality and VR, encompassing VR, AR, and MR \cite{morimoto2022xr}. VR, AR, and MR are three distinct technologies that enable immersive experiences in different ways. VR offers a fully digital environment, while AR overlays digital information onto the real world. MR, on the other hand, blends both VR and AR, allowing users to interact with virtual objects within their physical environment \cite{chuah2018and}. AR finds common use in mobile applications and smart glasses, enhancing users' perception of their surroundings. MR takes this a step further by anchoring digital objects to the real world and enabling users to interact with them as if they were tangible. This technology creates truly immersive experiences, where users can engage with virtual elements while remaining aware of and connected to the physical world. XR (Extended Reality) encompasses VR, AR, and MR, and it is finding increasing adoption across various industries, such as entertainment, gaming, education, healthcare, and business \cite{doolani2020review}.

\subsubsection{Avatars in Digital Spaces}
An avatar is a virtual representation of an individual or user within a digital space, often in the form of a 3D model. Avatars are used in video games, social media, and other virtual environments to represent a person's presence in that space. Avatars can be highly customizable, allowing users to express themselves in unique ways and interact with others in a virtual space \cite{9282914}.

\subsection{Architecture of Metaverse}
\begin{figure}
     \begin{subfigure}[b]{\textwidth}
         \includegraphics[width=.5\textwidth]{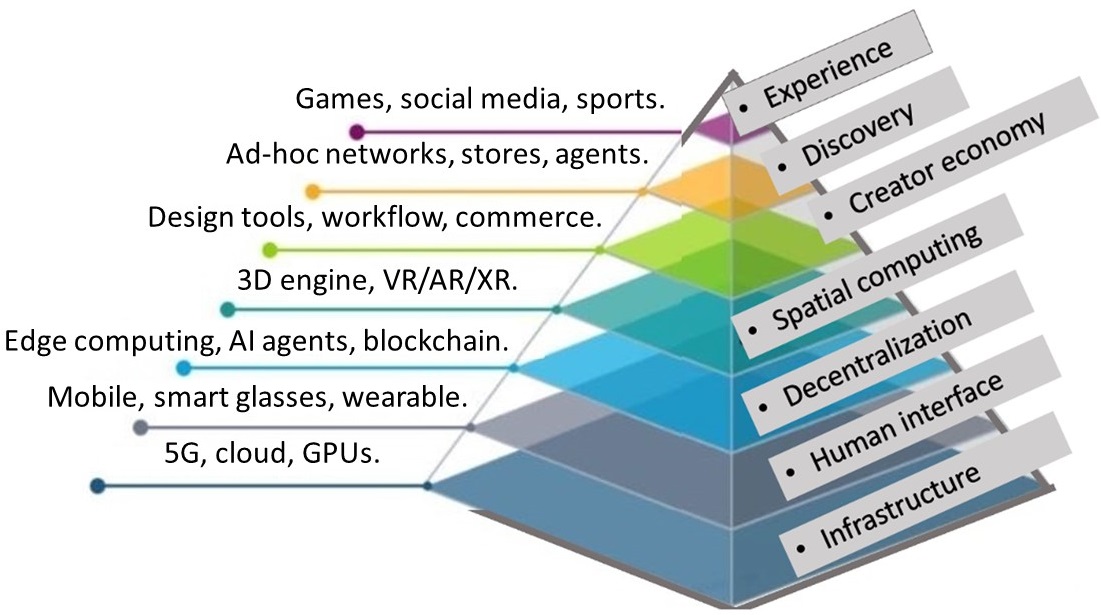}
     \end{subfigure}
     \begin{subfigure}[b]{\textwidth}
                  \includegraphics[width=.48\textwidth]{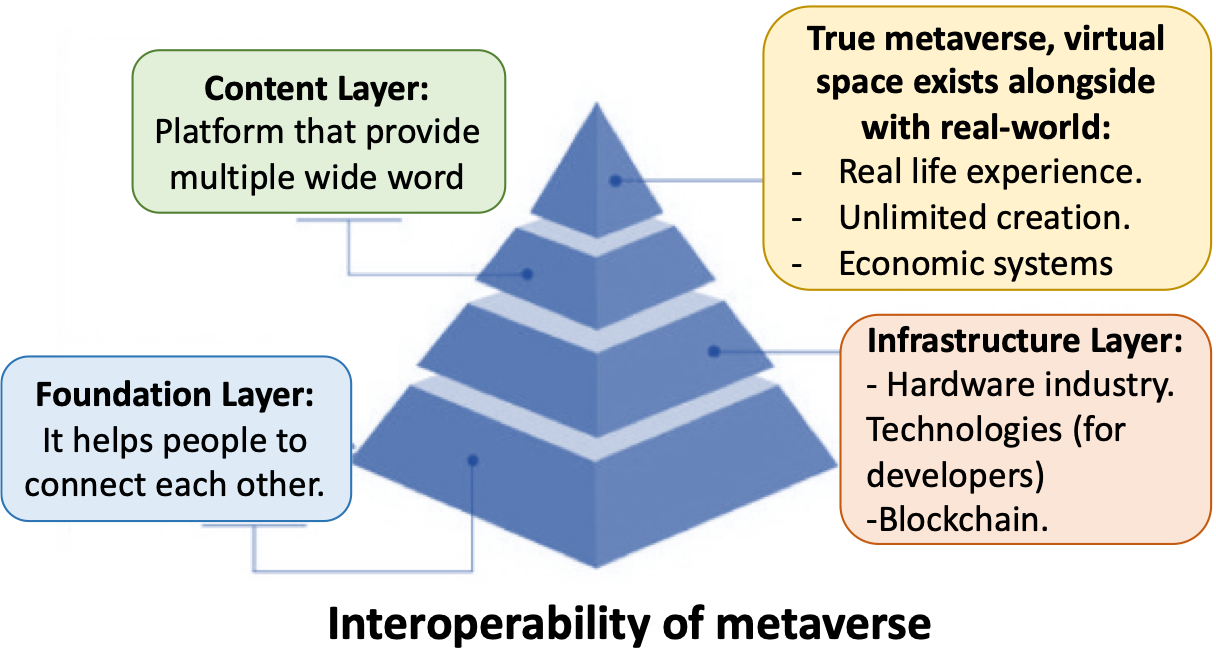}
     \end{subfigure}
        \caption{Identifying metaverse layers and issues related to metaverse interoperability}
        \label{Fig:MeteversArchitucture}
\end{figure}

Setiawan et al. \cite{9915136} discussed metaverse layers, which refer to the various components or building blocks that make up a metaverse, which is a virtual environment where users can interact with each other in a shared online space. The layers of a metaverse can be thought of as different aspects of the environment, each serving a specific purpose. The metaverse typically comprises several fundamental layers, including infrastructure, experience, discovery, creator economy, spatial computing, decentralization, and human interface, an illustration of different layers of the metaverse is shown in Figure \ref{Fig:MeteversArchitucture}.


\subsubsection{Metaverse Experience Layer}
The experience layer in the metaverse is the first and most vital layer, representing the 3D virtual world. It enables diverse interactions among users and with the environment, utilizing avatars for self-representation. Entertainment and social activities like games, virtual concerts, and meetups are common in this layer, aiming for a seamless and immersive experience that goes beyond real-life possibilities. Achieving interoperability relies on open standards, protocols, and APIs to facilitate data sharing across systems. Key challenges involve establishing universal standards and interfaces for platform communication and addressing identity and data portability concerns across diverse systems \cite{9915136}.

\subsubsection{Metaverse Discovery Layer}
The discovery layer is the second layer in the metaverse, responsible for connecting users with content, services, and other users. It includes search, recommendations, social networking, and community building tools. Users can access inbound experiences by actively searching for information or content, like browsing virtual stores for products. Outbound experiences involve information pushed to users, such as targeted advertising or event notifications. The discovery layer enables users to access a wide range of experiences, using features like search engines, recommendation algorithms, and social networks to navigate the metaverse \cite{9915136}.

\subsubsection{Metaverse Creator Economy Layer} 
The third layer of the metaverse is focused on empowering creators with the necessary tools to design and build their own content, without requiring programming knowledge. This includes 3D modeling, animation, and virtual world tools. The primary objective is to make it accessible for anyone to create content and share it with others in the metaverse. Furthermore, this layer provides a marketplace for creators to sell their content, which encourages more people to create content and ultimately helps to expand the metaverse, making it more vibrant and engaging for all users \cite{9915136}.

\subsubsection{Metaverse Spatial Computing Layer}
Spatial computing involves utilizing technology to generate an interactive and immersive digital environment that seamlessly merges with the physical world. This is achieved through the utilization of technologies such as VR, AR, cloud computing, sensors, and spatial mapping. Spatial computing is a critical aspect of the metaverse, enabling users to engage with the virtual world in a more organic and intuitive manner and facilitating more sophisticated and realistic virtual experiences \cite{9915136}.

\subsubsection{Metaverse Decentralization Layer}
The Decentralization Layer plays a critical role in providing a decentralized infrastructure that ensures security, privacy, and governance in the metaverse. This is achieved through the use of blockchain, which enables the creation of a decentralized network of nodes that can host and manage the metaverse data and assets. This layer ensures that metaverse data and assets are stored in a secure and tamper-proof manner, while firmly establishing ownership and control of assets. Additionally, it provides a mechanism for decentralized identity and authentication, enabling users to access and interact with the metaverse in a secure and private manner \cite{9915136}.

\subsubsection{Metaverse Human Interface Layer}
This layer involves the technology and devices such as VR headsets, AR glasses, haptic suits, motion sensors, and other input devices  that users interact with to access and experience the metaverse. Besides, this layer provides an intuitive and immersive user experience that enables users to navigate and interact with the metaverse seamlessly. Additionally, this layer will likely include features such as speech recognition, natural language processing (NLP), and other AI-based applications to enhance user experiences \cite{9915136}.

\subsubsection{Metaverse Infrastructure Layer}
The infrastructure layer forms the backbone of the metaverse, incorporating both hardware and software infrastructure for data storage, processing, and transmission across all layers. Communication protocols and network standards ensure seamless interoperability between virtual worlds and applications. Stability, scalability, and security are paramount in this layer, directly influencing the overall metaverse quality. Advanced technologies like cloud computing, edge computing, blockchain, and distributed storage systems are crucial for supporting the infrastructure. Additionally, cutting-edge wireless technologies (5G, 6G, Wi-Fi) and essential components like CPUs and batteries enable seamless integration and adaptability to the metaverse's dynamic nature \cite{9915136}.

\subsection{Interoperability of Metaverse}
Interoperability refers to the ability of different software systems or platforms to communicate and exchange information seamlessly. Interoperability in the metaverse can be achieved through open standards and protocols that allow different platforms and systems to communicate with each other. Interoperability allows users to have a seamless experience across different virtual worlds and platforms, which is essential for the growth and development of the metaverse. By enabling users to create and customize their avatars and virtual assets in one platform and use them in another without duplicative work, interoperability makes it easier for users to move between different virtual worlds and have a consistent experience. Cross-platform communication and transactions enable users to connect and collaborate with other users, regardless of the platform they are on. Finally, decentralized identity gives users more control over their identity and privacy, which is essential for building trust in the metaverse. Interoperability is a critical aspect of the metaverse, and it will be important for developers and stakeholders to work together to ensure that interoperability is maintained and expanded as the metaverse continues to grow \cite{li2023metaopera}.

\subsection{AI for Metaverse}

The metaverse can utilize pre-trained AI foundation models, extensive language models trained on vast text data. These models can be fine-tuned for specific tasks like text classification, question-answering, or sentiment analysis. Popular foundation models include Bidirectional Encoder Representations from Transformers (BERT), Generative Pretrained Transformer 3 (GPT-3), Robustly Optimized BERT Pretraining Approach (RoBERTa), and A Lite BERT (ALBERT). When integrating these models, careful consideration of potential risks and benefits is essential, along with implementing appropriate measures to protect sensitive information.

If we represent a foundation model mathematically as a function $F(x)$, where $x$ is an input and $F(x)$ is the model's prediction for the output, learned from a large dataset during pre-training, this function is defined by a set of parameters $\theta$. When using a foundation model for a specific task in AI-XR metaverse, a smaller labeled dataset for the task is employed to fine-tune the model, and update its parameters $\theta$. This can be represented mathematically as the following optimization problem:

\begin{equation}
 \text{minimize} \hspace{1mm} J(\theta) = 1/N \sum_{i=1}^n L(F(x_i; \theta), y_i)
\end{equation}

where $N$ is the number of examples in the labeled dataset, $x_i$ is the $i-th$ input, $y_i$ is its corresponding true label, $L(.)$ is a loss function measuring the difference between the model's prediction and the true label, and $\theta$ represent model's parameters. The goal is to find the optimal parameters $\theta$ that minimize the objective function $J(\theta)$ on the task-specific dataset. This fine-tuning process allows the foundation model to be adapted to the specific requirements of AI-XR metaverse, this makes it a valuable tool for improving the user experiences in virtual environments. 

In the next few subsections, we discuss how modern pre-trained transformer-based AI models, also known as \textit{foundation models} \cite{bommasani2021opportunities}, may be used in the context of the metaverse.

\begin{table*}[!h]
\caption{{Comparison and decomposition of foundation models, their potential applications in the metaverse, and challenges.} }
\label{tab:foundation-model}

\begin{tabular}{|l|p {1.5cm}|p {6cm}|p {6.5cm}|}

\hline
\textbf{Reference} & \textbf{Foundation Model} & \textbf{Potential Applications in the Metaverse}& \textbf{Challenges} \\ \hline
Firat et al. \cite{firat_2023}
&
 ChatGPT
&
Natural language interactions with virtual characters.
Virtual assistants and chatbots for immersive experiences. Language understanding and generation in virtual environments.
&
Ensuring coherent and contextually appropriate responses,  handling ambiguity, and understanding user intent in dynamic virtual worlds, real-time performance, and latency for interactive conversations.
\\ \hline
He et al. \cite{2023}
&
BERT
&
Natural language understanding in virtual environments, sentiment analysis, emotion detection in virtual interactions, text classification, and recommendation systems in the metaverse.
&
Training large-scale language models can be computationally expensive, fine-tuning for specific metaverse domains and contexts. Handling out-of-vocabulary words and new concepts in virtual environments.
\\ \hline
Delobelle et al. \cite{delobelle2020robbert}
&
RoBERTa
&
Language understanding and generation in VR scenarios, text summarization and content extraction for immersive experiences, sentiment analysis, and opinion mining in virtual social interactions.
&
Large storage and computational requirements during training, adapting the model to different metaverse platforms and languages, mitigating potential biases in training data and generated outputs.
\\ \hline
Wang et al. \cite{wang2020sentiment}
&
ALBERT
&
Efficient language understanding and generation in resource-constrained metaverse applications, cross-lingual understanding, and translation in virtual environments, content recommendation, and personalization for immersive experiences.
&
Balancing model size reduction with retaining language representation quality, fine-tuning for specific metaverse domains with limited labeled data, mitigating biases and fairness concerns in multilingual contexts
\\ \hline

\end{tabular}
\end{table*} 

\vspace{1mm}
\subsubsection{ChatGPT into AI-XR-enabled metaverse}
ChatGPT is a versatile AI tool that can be integrated into AI-XR-enabled metaverse environments to enhance user experiences. For instance, it can be utilized as a virtual assistant, dialogue generator, predictive text tool, and personalization engine. By integrating ChatGPT into virtual environments, it can provide real-time assistance, information, and support to users. Additionally, it can generate engaging and natural dialogue between users and virtual characters in VR and the metaverse. Moreover, ChatGPT can suggest text and responses to users in real time, making it easier to communicate and interact with other users within virtual environments. By leveraging ChatGPT's capabilities, users can enjoy more natural, intuitive, and personalized experiences within VR and the metaverse.

Firat et al. \cite{firat_2023} demonstrated that ChatGPT is a pre-trained model developed by OpenAI, designed to generate human-like text. Due to its training on an extensive corpus of text data, it can perform a broad spectrum of language-related tasks, including question-answering, text completion, summarization, translation, etc. This model like other large language models has raised some concerns regarding privacy. These concerns stem mainly from the fact that GPT-3 has been trained on a vast amount of personal and sensitive data, including private conversations, and emails, that may be contained in the text data utilized for model training. Additionally, the sheer size and power of GPT-3 raise questions about the potential misuse of the model to generate fake news, impersonate individuals online, or even manipulate public opinion. While OpenAI has implemented some privacy measures, such as redacting personally identifiable information, the risk of privacy breaches remains a concern. As with any technology, it is important to consider the potential risks and benefits and take steps to mitigate those risks.

\vspace{1mm}
\subsubsection{BERT into AI-XR-enabled metaverse}

He et al. \cite{2023} introduced BERT, a pre-trained language model developed by Google Research. BERT is designed for NLP tasks such as text classification and sentiment analysis, with fine-tuning capabilities for various NLP applications. Trained on a vast corpus of text data, BERT considers word context, improving its understanding of sentence meaning. BERT can process user-generated content, such as chat messages, social media posts, or reviews, to determine the sentiment expressed. This can be valuable for understanding user feedback, assessing the impact of virtual experiences, or even identifying potential issues such as cyberbullying within the metaverse. While BERT's application in the metaverse enhances natural language understanding, it raises privacy concerns as it may access sensitive information. Fine-tuning BERT on metaverse data enables personalized recommendations for virtual experiences, goods, and social connections, utilizing its contextual understanding for accurate and relevant suggestions. However, using BERT in the metaverse also raises ethical considerations about potential misuse, public opinion manipulation, decision influence, and privacy rights violation.

\vspace{1mm}
\subsubsection{RoBERTa into AI-XR-enabled Metaverses}
This AI model was developed by Facebook AI and it has been trained on large text data that includes scientific articles, books, and web pages, and has been shown to outperform BERT on a range of NLP tasks. Like BERT, RoBERTa can be fine-tuned for a variety of NLP tasks \cite{delobelle2020robbert}. However, it is also subject to the same privacy concerns as BERT, given that it was also trained on a large amount of text data that may contain personal and sensitive information. RoBERTa can also be applied to various tasks within the metaverse. RoBERTa model enhanced pretraining approach helps capture a deeper understanding of language semantics and improves the quality of generated dialog. By training RoBERTa on metaverse-related text, it can accurately recognize and classify named entities, enabling applications like virtual environment indexing, information retrieval, or data analysis within the metaverse.

\vspace{2mm}

\vspace{1mm}
\subsubsection{ALBERT into AI-XR-enabled metaverse} 
This model was developed by Google Research. ALBERT is used to address the computational and memory usage of the original BERT model by minimizing the number of parameters model and sharing parameters across multiple layers. ALBERT uses a factorized embedding parameterization, which reduces the number of parameters and allows the model to be trained more efficiently. Despite its reduced size, ALBERT is still a large language model that was trained on a massive amount of data and may contain personal and sensitive information \cite{wang2020sentiment}. As such, it is subject to the same privacy concerns as other language models and it is important to consider the potential risks and benefits and implement appropriate measures to protect sensitive information. ALBERT can be applied to text summarization tasks within the metaverse. This can involve summarizing long articles, user-generated content, or virtual environment descriptions into shorter and more concise summaries. By fine-tuning ALBERT on summarization datasets specific to the metaverse, it can generate accurate and relevant summaries. By training ALBERT on sentiment-labeled or emotion-labeled datasets, it can analyze user-generated content and determine sentiment or emotions expressed within virtual environments. This can help in understanding user feedback, monitoring community interactions, or improving virtual experiences.

\subsubsection{Other Applications of AI in metaverse}

AI applications in the metaverse, such as machine vision, generative modeling, speech processing, and NLP \cite{zhu2022metaaid,qayyum2022secure}, offer enhanced user experiences and foster new creativity and innovation. Generative modeling allows the creation of unique virtual environments, characters, and objects, facilitating efficient content generation based on existing data. Speech processing enables voice chat for more natural and immersive interactions between users, facilitating collaboration and socialization in virtual environments. NLP powers chatbots and virtual assistants that help users navigate the metaverse, perform tasks, and provide information, making it user-friendly, especially for newcomers. For a comparison and decomposition of foundation models, their potential metaverse applications, and challenges, refer to Table \ref{tab:foundation-model}.

\subsection{ML Security Issues}

Model performance is crucial in ML to achieve accuracy, generalizability, and user privacy preservation \cite{seif2020wireless}. Moreover, optimizing ML models for specific objectives can lead to unintended consequences and biases. For instance, predictive policing algorithms may unfairly target certain communities, resulting in biased treatment. Deep learning (DL) models, in particular, have been shown to be vulnerable to various adversarial attacks \cite{qayyum2020secure,qayyum2020securing} and poisoning attacks leading to security concerns in ML \cite{qayyum2022securing}. We discuss these attacks next.

\subsubsection{Adversarial ML Attacks}
Adversarial ML attacks involve malicious attempts to deceive or manipulate ML models using carefully crafted input data. For instance, attackers may manipulate a spam detection model to evade its algorithms or trick an image recognition model into misclassifying an altered image. Other security issues include model stealing, where attackers gain unauthorized access to a model and create a duplicate for exploitation, and data poisoning, where attackers manipulate training data to influence model predictions.

\subsubsection{Data Poisoning Attacks}
Data poisoning involves injecting malicious data into the training set, compromising model accuracy, or introducing bias. Model stealing enables the unauthorized use or reverse-engineering of a model, risking intellectual property exploitation. Additionally, ML models trained on sensitive data like medical or financial records can be susceptible to privacy violations if not properly secured. Attackers may infer sensitive information from model outputs or reverse-engineer the model.

\subsection{Secure Data Sharing}
To ensure privacy protection, it is essential to secure the transmission and processing of data, employing encryption and other security measures to prevent unauthorized access or disclosure. Data should only be shared with trusted parties who have a legitimate need and adhere to strict confidentiality agreements. Regular security audits and vulnerability assessments help identify and address potential privacy breaches proactively. In the metaverse, wearables like head-mounted displays (HMDs) can collect significant amounts of personally identifiable information \cite{wei2020ldp}. To maintain confidentiality, this data should be transmitted using a secure communication channel \cite{qayyum2023can}. However, adversaries can still intercept data through eavesdropping, and advanced techniques like differential and inference attacks can compromise user privacy, including location tracking and access to raw data \cite{wang2022survey}. Storing large amounts of private and sensitive information, such as user profiles, on servers or edge devices increases the risk of privacy breaches. Hackers can exploit various attack vectors, exposing user privacy through frequent and diverse attacks. Furthermore, storing data on cloud or edge devices may be vulnerable to distributed denial-of-service (DDoS) attacks, as noted in previous research \cite{bertino2017botnets}.

\section{Challenges in Metaverse}
\label{sec:Challenges}

\subsection{Metaverse Related Security and Safety Issues}

\subsubsection{Single Point of Failure}
In metaverse systems that utilize a centralized architecture, such as cloud-based systems, there is a potential risk of a single point of failure. Although this type of architecture offers advantages such as simplified user/avatar management and operational cost savings, it can be susceptible to a single point of failure due to physical root server damage and distributed DDoS attacks. Furthermore, this architecture may present challenges regarding trust and transparency in the secure exchange of virtual assets, currencies, and goods among different virtual environments within the metaverse \cite{wang2021blockchain}.

\subsubsection{Denial-of-Service (DDoS)}
With the integration of a noteworthy quantity of wearable devices, the metaverse faces the risk of potential compromise by malicious actors who could exploit these devices to create a botnet, such as the Mirai botnet \cite{rasool2022security}, and launch distributed DDoS attacks. Such attacks can cause network outages and service disruptions by inundating the server with a vast volume of traffic within a period. Moreover, due to the communication and storage constraints of blockchain technology, certain non-fungible token (NFT) functions that may be executed on off-chain systems. In such instances, bad actors could execute DDoS attacks to disrupt the NFT system's services, leading to service unavailability.

\subsubsection{Sybil Attacks}
Sybil attacks pose a significant threat where malicious actors can create numerous fake or stolen identities to gain an excessive amount of influence on metaverse services, such as blockchain consensus, reputation service, and voting-based service in digital governance \cite{liao2021digital}. This can undermine the system's efficacy and even permit attackers to seize control of the metaverse network. For example, by producing an adequate number of Sybil identities to out-vote authentic nodes, opponents can reject the delivery or receipt of particular blocks, thereby obstructing other nodes from accessing the blockchain network in the metaverse.

\subsubsection{Fraud Risks}
The metaverse presents significant fraud risks during the creation of user-generated content (UGC) and the trading of virtual objects among various stakeholders. These risks include incidents of repudiation and refusal to pay, which can compromise the metaverse's integrity. It is also essential to ensure the legitimacy and reliability of digital copies when using digital twin technologies to create virtual objects. Dishonest users or avatars could buy virtual items or UGC and unlawfully sell digital copies to unsuspecting parties for monetary gain. Moreover, opponents may capitalize on weaknesses in metaverse systems to commit fraudulent activities and undermine the credibility of services. An instance of this form of fraud is the Paraluni metaverse initiative, which is built on Binance Smart Chain, and suffered a loss of over 1.7 million in 2022 owing to a reentrancy flaw in its smart contracts \cite{wang2022survey}.

\subsubsection{Physical World and Human Society Threats}
As an extension of the cyber-physical-social system (CPSS), the metaverse represents a highly interconnected network of physical and cyber systems. Actions and events that occur within the virtual world can have significant impacts on the physical world and human society, making the threats in virtual worlds potentially dangerous. These threats can lead to severe consequences such as physical infrastructure damage, personal safety risks, and societal implications \cite{zhou2019cyber}.

\subsubsection{Personal Safety}
Adversaries can take advantage of weaknesses in wearable devices, XR helmets, and indoor sensors to access personal information and track the real-time location of users, as stated in \cite{casey2019immersive}. This type of attack can enable criminal activities such as burglary. Moreover, attackers could potentially manipulate a VR device to deceive individuals and lead them into harmful physical situations, which could result in serious physical harm.

Metaverse sensors present several privacy threats to users. These sensors can track a user's location, movements, and behavior, creating a profile of their activity over time, which could be used for targeted advertising or tracking user behavior for other purposes. Some sensors include microphones and cameras, which could capture audio and video of users without their consent, potentially leading to a significant privacy violation, especially if the data is shared with third parties. In addition, some metaverse sensors may collect biometric data, such as heart rate or facial expressions, which can be used to infer personal information about the user, and metadata that can be used to identify and track users.

\begin{figure}[!h]
\centering
\includegraphics[width=8cm]{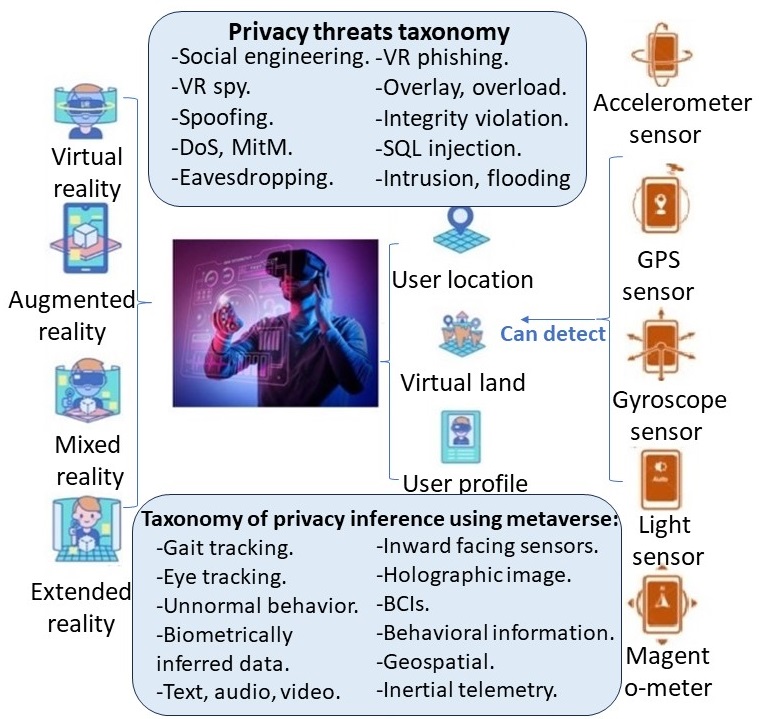}
\caption{Taxonomy of privacy threats opened up by metaverse sensors.}
\label{Fig:MetaverseSensorsThreats}
\end{figure}

Figure \ref{Fig:MetaverseSensorsThreats} highlights a taxonomy of privacy inferences and threats in the context of the metaverse and the need for robust privacy protections and systems to ensure the safe and secure use of these platforms.

\subsection{Metaverse Related Privacy Issues}
The metaverse gained attention during the COVID-19 pandemic and Facebook's re-branding as Meta, sparking debates about its potential as the future of work and play. The pandemic emphasized the need for alternative socialization methods, leading to increased interest in virtual environments like VRChat, Second Life, and Minecraft. Businesses also embraced virtual platforms for remote work and events. The metaverse's potential for innovative solutions, like healthcare simulations and virtual classrooms, came to the forefront. However, concerns about data security and physical-layer security risks in wireless communication remain \cite{cheng2023understanding, parker2023towards, falchuk2018social}. The metaverse's accumulation of vast amounts of sensitive data raises concerns about the potential for serious crimes, as outlined by Falchuk et al. \cite{falchuk2018social}. Furthermore, as the metaverse expands to integrate existing wireless communication technologies, traditional threats like physical-layer security risks to communication networks gain increased potency.

Figure \ref{Fig:MetaverseSensorsThreats} illustrates how metaverse sensors have the ability to collect extensive user data, including physical location, movement patterns, biometric data, and other confidential information. This data can be exploited in various ways, such as targeted advertising or identity theft. Moreover, metaverse sensors are vulnerable to hacking and other cyber attacks, potentially exposing user data to unauthorized parties.

\subsubsection{Accelerometer Sensor Privacy Threats}

Accelerometer sensors measure acceleration, tilt, and vibration of devices or objects \cite{faisal2019review}. Integrated into mobile and VR devices, they detect changes in device orientation, enabling motion-based features like gaming or fitness tracking. However, using accelerometer sensors to collect data on a user's physical movements raises privacy concerns. This data can reveal private information, including health status, location, and activities, and may be exploited for profiling or targeted advertising \cite{anand2019spearphone}. Unauthorized access to sensitive information is also a risk if the sensor is compromised \cite{anand2019spearphone}.

\subsubsection{Gyroscope Sensor Privacy Threats}

The gyroscope sensor, commonly found in smartphones, tablets, and VR devices, measures angular velocity or rotation, enhancing motion sensing accuracy in gaming and navigation. Comprising a micro-electromechanical system (MEMS), it uses Coriolis force during device rotation to provide orientation and movement information. When combined with accelerometers and magnetometers, gyroscope sensors offer comprehensive motion and orientation data, enabling features like screen rotation, gesture recognition, and AR applications. Gyroscope sensors alone do not pose serious privacy threats as they don't directly collect personal data. However, when combined with sensors like accelerometers or GPS, personal information such as location, movements, and activities can be potentially inferred. For example, with an accelerometer, a gyroscope sensor can track a user's movement and orientation in virtual or physical environments, potentially revealing sensitive information. Moreover, along with data sources like browsing history or search queries, the collected gyroscope data can be potentially exploited for targeted advertising or profiling \cite{9431220}.

\subsubsection{Magnetometer Sensor Privacy Threats}
A magnetometer sensor finds widespread usage in navigation systems, scientific research, mobile devices, and VR devices for measuring the strength and direction of magnetic fields. This sensor detects variations in magnetic fields as it moves and can identify both static and dynamic magnetic fields, offering data on their direction and intensity. When integrated with other sensors like accelerometers and gyroscopes, a magnetometer sensor can provide a more holistic understanding of a user's location and orientation. However, the amalgamation of this information could potentially be misused for targeted advertising, profiling, or other privacy infringements \cite{9431220}.

\subsubsection{Proximity Sensor Security Threats}

A proximity sensor detects nearby objects without physical contact \cite{9431220}. These sensors emit electromagnetic fields or radiation and measure changes caused by an object's presence or absence. In mobile devices, they turn off the display when the user holds the device to their ear during a call, conserving power. Proximity sensors are widely used in industrial automation, robotics, and automotive systems. When combined with cameras or microphones, they can collect more detailed data on a user's behavior or environment, potentially raising tracking and privacy concerns. Additionally, proximity spoofing attacks can exploit the sensor's readings, causing unintended actions. Ensuring proximity sensor security is essential to prevent vulnerabilities to such attacks \cite{9736583}.

\subsubsection{Light Sensor Privacy Threats}
A light sensor is a device that detects the intensity of light in an environment and converts it into an electrical signal for utilization by a device or system. In mobile devices, light sensors are commonly employed to optimize screen brightness, extend battery life, and enable features like automatic camera flash \cite{10.1145/3359789.3359840}. They are also utilized in various other applications, including automated lighting systems, security systems, and environmental monitoring. However, the use of a light sensor to trigger functions such as camera flash raises concerns about potential compromises to user privacy. It can enable the tracking of user activities or even capture images or videos without their consent \cite{9372295}.

\subsubsection{Eye Tracking Sensor Privacy Threats}
An eye-tracking sensor is a technology that monitors a user's eye movements and gaze direction. It utilizes infrared light emission and captures the reflection of that light from the user's eyes using a camera or another sensor. Eye-tracking sensors find numerous applications in user experience testing, gaming, and assistive technologies for individuals with physical disabilities. They can also enhance the accuracy and precision of augmented and VR systems \cite{10.1145/3359626}.

\subsubsection{Inside-Out Tracking Cameras Privacy Threats}
Inside-out tracking cameras are sensors utilized in VR and AR devices to track the user's movements and position in 3D space without external sensors. Positioned on the device's front, they employ computer vision technology for real-time tracking, offering flexibility and convenience in movement without being tethered to a fixed location. These cameras have diverse applications in gaming, education, architecture, and design. However, it's crucial to consider security risks, such as hacking or the collection of sensitive user data, associated with these sensors. Working with trusted devices and implementing appropriate security measures is essential. Inside-out tracking cameras use the VR device's inward-facing cameras, reducing the likelihood of capturing sensitive information. Nevertheless, hackers could still potentially access the camera for spying or maliciously tracking user movements. To mitigate these risks, users should use trusted VR devices (updated with security patches), secure their home network, and use strong unique passwords to prevent unauthorized access to their devices.

\subsubsection{Pervasive Data Gathering}
Pervasive data collection could occur through various means, such as tracking user movements and actions within the virtual environment, monitoring communication between users, and gathering information about user preferences and behaviors. Through the use of advanced XR and HCI (human-computer interaction) technologies, various forms of user data can be collected, such as eye/hand movements, speech, biometric features, and even brain wave patterns \cite{shang2020arspy}. Additionally, these technologies can enable the analysis of physical movements and user attributes, and facilitate user tracking. For example, the Oculus helmet's motion sensors and built-in cameras can track the user's head direction and movement, create a virtual representation of the user's surroundings and monitor their position and environment with high accuracy while using platforms like Roblox. However, if this device were to be hacked by malicious actors, the vast amount of sensitive data collected could be used to commit serious crimes.

\subsection{Trustworthy AI and AI-Related Privacy Issues}

Addressing trustworthiness in AI involves tackling bias, fairness, transparency, explainability, privacy, and AI governance. To establish trustworthy systems, minimizing bias and ensuring fairness is crucial to avoid perpetuating societal biases and discrimination \cite{rasheed2022explainable}. Transparency and explainability in complex AI systems, especially those using large datasets and intricate algorithms, are essential to build trust by enabling understanding of the decision-making process. Robust privacy and security measures are vital due to the sensitive personal data AI systems access, preventing compromise or misuse. Trust in AI can be reinforced by establishing clear accountability for unintended consequences or errors caused by AI systems. Effective governance and regulation are necessary to foster ethical and responsible AI development and use, as trust in AI governance directly impacts overall trust in AI.

\begin{table*}[!h]
\centering
\caption{{Machine Learning (ML) Based Solutions for Mitigating AI-XR Metaverse Attacks.} }
\label{tab:MlApproaches2protectMetavers}
\scalebox{0.95}{
\begin{tabular}{|l|p {4.5cm}|p {2.5cm}|p {8cm}|p |}

\hline
\textbf{Reference} & \textbf{Functionalities} & \textbf{ML Algorithm}& \textbf{Descriptions - Remarks} \\ \hline
\cite{hussain2020machine}
&
Intrusion attempts,
unusual network activity,
malware outbreaks,
preventing security breaches. 
&
ML-based anomaly detection methods.
&
Identify unusual patterns or behavior that deviate from normal behavior and can be applied using statistical algorithms, ML techniques, or a combination of both.
 \\ \hline
\cite{8302863}
&
Intrusion detection, malware detection, and network traffic analysis.
&
Deep learning
&
Analyze large amounts of data, identify patterns and relationships, and make predictions or decisions based on that information.
\\ \hline
\cite{viegas2020reliable}
&
Intrusion detection, malware classification, network traffic analysis
&
Decision trees
&
 The final result is a tree that can be used to make predictions. 
\\ \hline
\cite{alam2013random}
&
Intrusion detection, malware classification, and network traffic analysis.
&
Random forests
&
 Combines the predictions of all trees to make a final prediction. This reduces the over-fitting issue in single decision tree models and improves the accuracy and robustness.
\\ \hline
\cite{8300282},  \cite{2019}, \cite{9695995}
&
Intrusion detection, malware classification, network traffic analysis, distributed DoS attack.
&
Support vector machines (SVMs)
&
This algorithm is applicable in classification and regression problems. Its functioning involves the identification of a hyperplane that maximizes the margin between classes within the training data.
\\ \hline
\cite{7047716}
&
Intrusion detection, malware classification, and network traffic analysis. 
&
Naive Bayes classifier.
&
It is used for classification problems. 
The algorithm makes predictions based on the Bayes theorem.
\\ \hline
\cite{2022}
&
Intrusion detection, malware classification, network traffic analysis, distributed DoS attack.
&
K-nearest neighbor (KNN).
&
This algorithm operates by identifying the k nearest data points and subsequently deriving a prediction by considering either the majority class or average value of these neighboring data points.
\\ \hline
\cite{HADDADPAJOUH201888}
&
Intrusion detection, network traffic control, adaptive security policies.
&
Reinforcement learning
&
Used to solve sequential decision-making problems.
The algorithm learns from trial-and-error interactions with the environment.
\\ \hline
 \cite{KARBAB2018S48}, \cite{8377943}
&
Intrusion detection, malware classification, and network traffic analysis. 
&
Neural networks (NN)
&
The neurons are adjusted during training to minimize the prediction error. NNs are highly flexible and can handle complex non-linear relationships between inputs and outputs.
\\ \hline
\cite{7060117}
&
Intrusion detection, malware classification, and network traffic analysis.
&
Logistic regression
&
It is a statistical method used for binary classification problems, and it can also be extended to multi-class classification.
\\ \hline
\end{tabular}}
\end{table*}

\begin{figure}[!h]
\centering
\includegraphics[width=7cm]{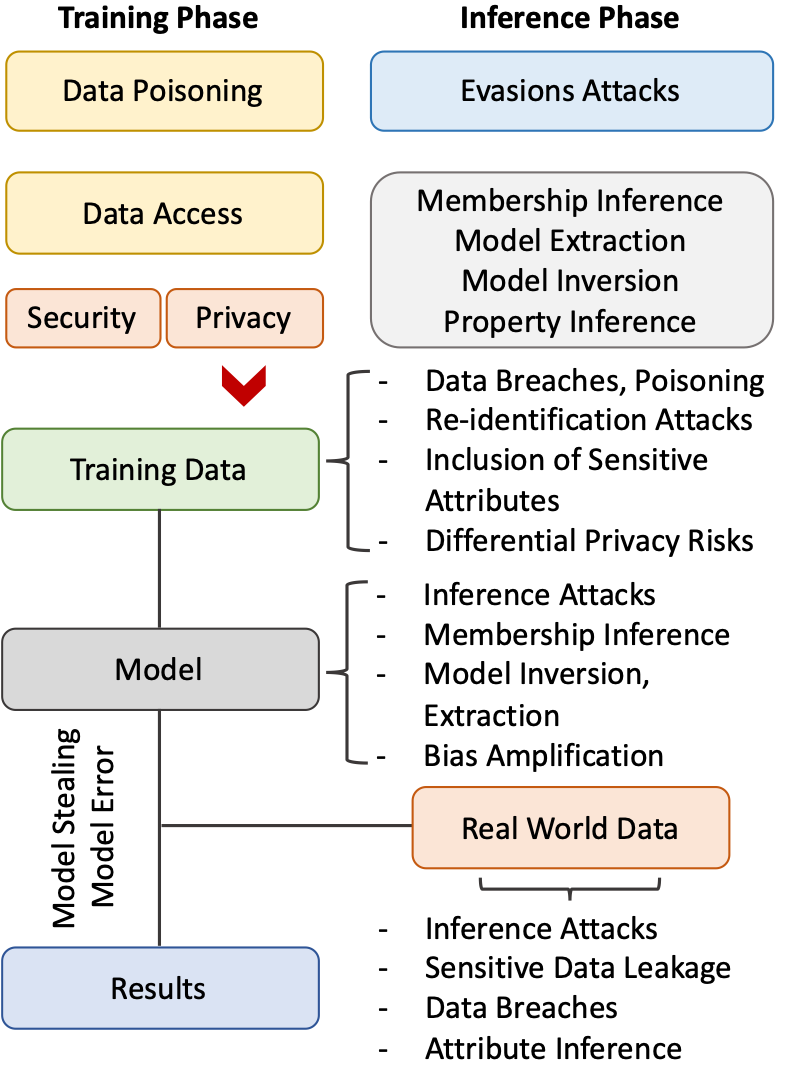}
\caption{Identifying privacy threats at AI applications.}
\label{Fig:AI-threats}
\end{figure}

ML has emerged as a powerful tool in various domains, including computer vision, NLP, and robotics, enabling the solution of complex problems. However, the increasing deployment of ML models in sensitive applications necessitates careful consideration of the potential privacy risks they pose. Figure \ref{Fig:AI-threats} provides a comprehensive overview and taxonomy of AI-related privacy concerns, encompassing issues such as bias, lack of transparency, and cybersecurity risks. 

Specific privacy threats like virtual stalking, virtual theft, identity theft, and behavioral tracking have been identified in the metaverse context. Privacy-preserving techniques are employed to protect users and applications in the metaverse; however, concerns arise due to the collection and sharing of personal data, necessitating awareness of data usage and recipients. Data privacy involves safeguarding sensitive information associated with ML models, including training data, model parameters, and predictions, as these models may inadvertently reveal personal details such as medical history, financial status, or personal preferences. Techniques like differential privacy, secure multi-party computation, and homomorphic encryption maintain model privacy, preserving the confidentiality of data and model predictions. Additionally, data anonymization, data minimization, and access control policies mitigate data breaches and unauthorized access to sensitive information, ensuring comprehensive privacy protection.

Various ML and DL approaches can be employed to address privacy concerns in AI-XR-enabled metaverse applications, as summarized in Table \ref{tab:MlApproaches2protectMetavers}, providing a comprehensive overview of different strategies to protect AI-XR metaverse applications from potential attacks.

\begin{table*}[!h]
\caption{{Summary of Relevant Software Libraries and Technologies for Metaverse Privacy.} }
\label{tab:software_libraries}

\begin{tabular}{|l|p {1cm}|p {5cm}|p {8cm}|}

\hline
\textbf{Reference} & \textbf{Year} & \textbf{Software Libraries/Technologies}& \textbf{Insights for Privacy in the Metaverse} \\ \hline
Smethurst et al. \cite{smethurst2023digital}
&
 2023
&
Self-sovereign identity (SSI) frameworks (e.g., Sovrin, uPort, Veres One)
&
Enable users to control and manage their own digital identities, prioritizing privacy by avoiding reliance on centralized authorities or third-party platforms.
\\ \hline
Zichichi et al. \cite{zichichi2023protecting}
&
2023 
&
Decentralized Identifiers (DIDs) and Verifiable Credentials (VCs) specifications
&
Provide a privacy-enhancing foundation for identity management in the metaverse, enabling secure and privacy-preserving authentication and data sharing between different components.
\\ \hline
Wan et al. \cite{wan2023web3}
&
2023 
&
Blockchain platforms (e.g., Ethereum, EOS, Polkadot)
&
Offer decentralized, immutable, and transparent systems for recording and verifying transactions. However, privacy concerns arise due to the public nature of the ledger.
\\ \hline
Xiao et al. \cite{xiao2023blockchain}
&
2023
&
Privacy-focused blockchain technologies (e.g., Monero, Zcash, Mimblewimble)
&
Implement cryptographic techniques like zero-knowledge proofs and ring signatures to enhance privacy and anonymity for transactions in the metaverse. Can be integrated to protect sensitive user information.
\\ \hline

\end{tabular}
\end{table*}

Amich et al. \cite{amich2021explanation} presented that an attacker manipulates the input to a trained ML model to induce the model to make an error. For example, an attacker might add noise to an image to cause an image classifier to misclassify the image, and this will threaten privacy. Lin et al. \cite{lin2021ml}, presented that an attacker might also add a small number of carefully crafted examples to a dataset used to train a classifier, causing the classifier to misclassify a specific type of object. Elsayed et al. \cite{elsayed2020ddosnet}, presented that an attacker may use techniques such as gradient-based optimization to create adversarial instances or samples that are almost indistinguishable from the original input, threatening the ML model privacy.

Furthermore, it is essential to protect the models themselves, the data they are trained on, and the infrastructures they run on. Adversarial examples can also be used to attack DL models. Data skewing is a phenomenon in ML where data distribution in a dataset is highly imbalanced. When a dataset is highly imbalanced, ML models tend to perform poorly in the minority class. This is because the model may be biased towards the majority class and have difficulty learning the patterns in the minority class \cite{lim2020federated}. Data imbalance, or class imbalance, is a prevalent issue in ML. It arises when the distribution of classes in a dataset is uneven, with one or more classes having significantly fewer instances than the others. This can be problematic since ML models usually prefer the majority class, resulting in inadequate performance on the minority classes \cite{ma2020safeguarding,yang2019federated}.
ML and DL models can be vulnerable to privacy concerns in several ways. For example, model inversion attacks, membership inference attacks, overfitting, and data leakage. If a model's training data contains sensitive information that is not adequately anonymized or protected, it could be exposed through its outputs, as shown in Table \ref{tab:MLVulnerabilities}.

\begin{table*}[]
\centering
\caption{Types of ML Vulnerabilities}
\label{tab:MLVulnerabilities}
\scalebox{0.95}{
\begin{tabular}{|l|p {3cm}|p {7cm}|p {6cm}|p |}
\hline
\textbf{Reference} & \textbf{Vulnerabilities} & \textbf{Descriptions - Remarks} & \textbf{Implications }\\ \hline
  \cite{Ying_2019}, \cite{9023664}
&
Inference Attacks 
&
involve inferring sensitive information about individuals by analyzing the outputs or predictions of an ML model.
&
Attackers may deduct private details that were not intended to be disclosed.
 \\ \hline
 \cite{10.1145/3457607}
&
 Bias 
&
Refers to the systemic error in a model that leads it to consistently make incorrect predictions for certain groups of individuals.
&
Discrimination and unfair treatment, limited representation,
lack of accountability and transparency
\\ \hline 
\cite{10.1007/978-3-030-90019-9_11} 
&
Model Inversion 
&
 Attempt to reconstruct training data or inputs based on the outputs or predictions of an ML model
&
Attackers can repeatedly query the model and use optimization techniques to approximate the original data.
\\ \hline 
 \cite{Collinse048008}
&
Explanation problems 
&
Many ML models, are highly complex making it difficult to understand their decision-making process. 
&
Lack of trust and acceptance, legal and ethical concerns, and transparency and accountability challenges.
\\ \hline 
 \cite{8677282}, \cite{10.1145/3436755}
&
Data Leakage
&
This can occur when the model learns to rely on irrelevant or sensitive features.
&
lack of informed consent and transparency.
\\ \hline 
\cite{pmlr-v139-koh21a}
&
 Differential Privacy Risks 
&
 Employed to protect individual privacy in ML training data.
 &
Risk of under-protecting or overprotecting the data
\\ \hline 

\end{tabular}}
\end{table*}

\subsection{General Challenges}
The metaverse development presents numerous technical and social challenges that should be addressed to create successful metaverses. Achieving a seamless and immersive metaverse will necessitate substantial investments in computing power, data storage, and network infrastructure. To support real-time interactions among a large number of users and digital objects, the metaverse will require considerable computing resources. Researchers like Jovanović et al. \cite{jovanovic2022vortex} emphasizes the reliance on AR and VR technologies, while others, as highlighted in \cite{cha2022performance}, focus on creating a more interconnected and immersive reality. Facilitating data exchange across AR and VR platforms is crucial for solving real-world problems and driving innovation in the metaverse.

Moreover, the metaverse will need extensive storage capacity to handle vast amounts of data related to user interactions, object properties, and other aspects of the virtual environment. Providing high-speed, low-latency network connectivity is also essential for enabling real-time interactions between users and digital objects. Meeting these infrastructure requirements will likely demand significant investments in new technologies and infrastructure. Deploying advanced data storage technologies capable of managing the massive data generated by user interactions and establishing high-speed, low-latency network infrastructure, such as 5G wireless networks, will be vital for delivering a seamless, real-time experience to users.

Addressing these infrastructure challenges necessitates collaboration among technological companies, governments, and other stakeholders. Governments may need to invest in new infrastructure projects to support the metaverse, while technology companies should develop innovative technologies and platforms. Ultimately, the successful development of the metaverse hinges on substantial infrastructure investments that can underpin the creation of a seamless and immersive virtual environment.

\section{Solutions for Assuring Privacy in AI-XR metaverse} 
\label{sec:Solutions}

\begin{table*}[]
\centering
\caption{Overview of Different Privacy Attacks on ML Models.}
\label{tab:Summarize the different approaches aimed at achieving privacy attacks on ML models}
\scalebox{0.95}{
\begin{tabular}{|p {1cm}|p {2cm}|p {2cm}|p {2.5cm}|p {3cm}|p {6cm}|}
\hline
\textbf{Reference} & \textbf{Privacy Attack} & \textbf{Threat Model} & \textbf{Application Attacked }& \textbf{Attack Success}& \textbf{Limitations}\\ \hline
\cite{hu2022membership}  
&
Membership Inference 
&
Adversarial Individual
&
Various ML Applications
&
Determine membership information
&
Limited success against models with strong defenses or noise injection.
\\ \hline 
\cite{zhang2020secret}
&
Model Inversion 
&
Adversarial Individual
&
Image Recognition
&
Reconstruction of sensitive data
&
Highly dependent on the model architecture and data availability
\\ \hline 
\cite{al2019privacy}
&
Reconstruction Attack
&
Adversarial Individual
&
Health Records
&
Reconstruct sensitive training data
&
Requires access to model parameters or gradients and can be challenging for complex models
\\ \hline
\cite{yuan2021current}
&
Adversarial Examples
&
Adversarial Individual
&
Image Classification
&
Mislead the model's predictions
&
Vulnerable to detection and defenses such as adversarial training
\\ \hline
\cite{miao2021machine}
&
Model Stealing
&
Adversarial Individual
&
Model as a Service/API
&
Extract the ML model or approximation
&
Success highly depends on the model's architecture and defenses implemented by the service provider
\\ \hline
\cite{liakos2020conventional}
&
Trojan Attack
&
Insider Threat
&
Various ML Applications
&
Trigger malicious behavior or data leaks
&
Requires access to the training process and model parameters
\\ \hline
\cite{zhang2022survey}
&
Attribute Inference
&
Adversarial Individual
&
Recommender Systems
&
Infer sensitive attributes or features
&
Limited success against models with strong privacy protection mechanisms
\\ \hline
\cite{miao2021machine}
&
Model Extraction
&
Adversarial Individual
&
Black-Box ML Models
&
Extract internal model information
&
Approximation may result in lower model performance and limited accuracy
\\ \hline
\end{tabular}}
\end{table*}

There are several software libraries and technologies that can be utilized in the metaverse to address privacy concerns. One notable example is Self-sovereign identity (SSI) frameworks, including Sovrin, uPort, and Veres One. These frameworks empower users to have control over and manage their digital identities. These frameworks prioritize privacy by eliminating the need for centralized authorities or third-party platforms. In the context of the metaverse, the Decentralized Identifiers (DIDs) and Verifiable Credentials (VCs) specifications establish a privacy-focused foundation for identity management. These specifications make secure and privacy-preserving authentication and data sharing between different metaverse components possible.

Blockchain platforms like Ethereum, EOS, and Polkadot offer decentralized, immutable, and transparent systems for recording and verifying transactions in the metaverse. However, the public nature of the blockchain ledger raises privacy concerns. To address this issue, privacy-focused blockchain technologies can be used such as Monero, Zcash, and Mimblewimble--employ cryptographic techniques like zero-knowledge proofs and ring signatures. These techniques enhance privacy and anonymity for transactions within the metaverse. Integrating these privacy features into metaverse systems helps safeguard sensitive user information.

Table \ref{tab:software_libraries} provides a summary highlighting the key insights of the mentioned software libraries and technologies regarding privacy aspects in the metaverse.

\subsection{AI-XR metaverse privacy-enhancing technologies}

Privacy-enhancing technologies (PETs) focus on developing techniques and tools that can be used to protect an individual's data and privacy in various contexts, including the use of AI and XR technologies. PETs are designed to mitigate the risks associated with data collection, storage, and processing and can also include techniques such as encryption, anonymization, and differential privacy. These studies aim to identify the most effective PETs for specific use cases and contexts and evaluate their effectiveness in protecting AI-XR metaverses' privacy. Striking the right balance between privacy, security, and usability while implementing technical safeguards and socio-technical considerations is vital to ensure a responsible and user-centric development of the metaverse.

Technical measures, including strong encryption, firewalls, intrusion detection systems, anti-virus software, and secure transactions, play a crucial role in safeguarding sensitive information shared within the metaverse. Park et al. \cite{park2022Metaverse} emphasized the importance of considering factors such as interpretability, privacy, societal function, and ethics to develop sustainable metaverse applications. Striking the right balance between privacy, security, and usability is essential. Implementing two-factor authentication for avatars and encrypting transmitted data can contribute to protecting user privacy. However, it is crucial to avoid excessive restrictions or censorship that may hinder the metaverse's potential.

Protecting user privacy and content is crucial for the safe and responsible development of the metaverse. The creation and sharing of digital assets such as 3D models, textures, and scripts by users can lead to intellectual property issues. To address this, metaverse platforms should provide mechanisms for content creators to protect their assets, such as digital rights management (DRM) and the use of blockchain to create a secure and transparent record of ownership \cite{lv2022Metaverse, maksymyuk2022blockchain, mishra2022contribution, luo2021novel}. The metaverse is also vulnerable to cyber-attacks and malicious activities like hacking, identity theft, fraud, deception, and exposure to harmful content \cite{di2021Metaverse, hilt2020caught, mackenzie2022criminology}. To ensure user privacy, metaverse platforms, and creators must prioritize privacy by implementing robust measures, protecting user-generated content, and being transparent about data collection and usage \cite{carlini2021poisoning}. Furthermore, using privacy-preserving methods in developing ML models, such as differential privacy, secure multi-party computation, and HE, is also crucial \cite{carlini2021poisoning}.

Ensuring the privacy of user-generated private data in the metaverse necessitates metaverse platforms to offer robust mechanisms for creators to safeguard their digital assets, including 3D models, textures, and scripts, which may hold significant value and be protected by intellectual property rights. Digital rights management (DRM) implementation and the utilization of blockchain technology for secure ownership records \cite{lv2022Metaverse, maksymyuk2022blockchain, mishra2022contribution, luo2021novel} are vital in this regard.

\begin{figure}[!t]
\centering
\includegraphics[width=8cm]{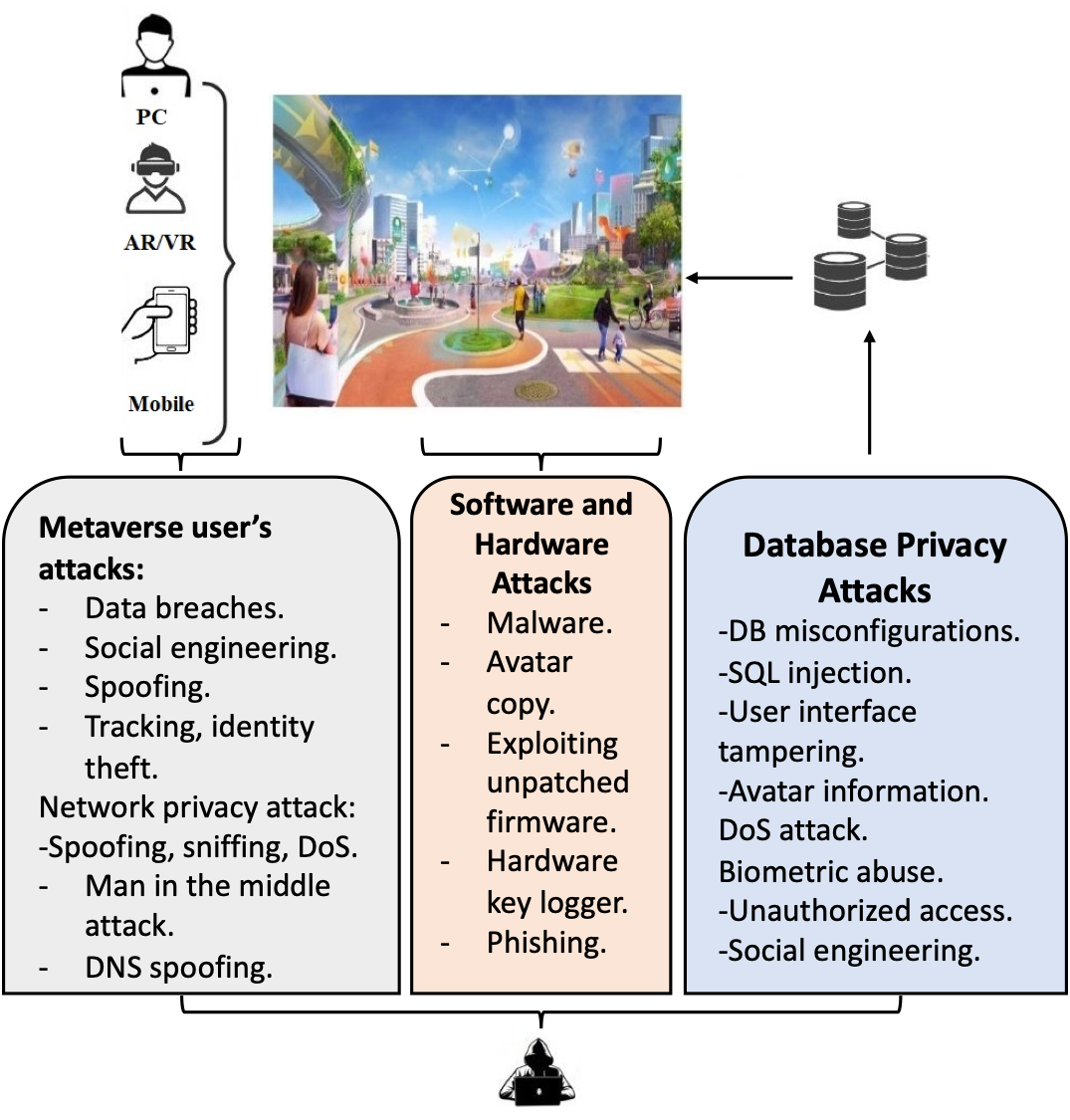}
\caption{Exploring privacy risks in metaverse platforms.}
\label{Fig:MetaverseAndMLthreats}
\end{figure}

Maksymyuk et al. \cite{maksymyuk2022blockchain} argued that blockchain technology is a key enabler for the metaverse system due to its decentralized nature, which is essential for synchronizing physical and VR. The distributed ledger technology offered by blockchain can provide a secure and transparent method for content sharing in the metaverse, as it is validated by a consensus algorithm that ensures the integrity of the data stored on the ledger. This technology provides a secure and tamper-proof environment for storing and managing data in the metaverse.

Mishra et al. \cite{mishra2022contribution} presented that incorporating blockchain into the metaverse can offer several benefits, particularly regarding data privacy. The 
decentralized nature of blockchain provides a secure and transparent way to store and manage user data. The authentication and access control methods and consensus processes ensure that only authorized users can access the data, thus protecting user privacy. Luo et al. \cite{luo2021novel} suggested that blockchain technology also helps maintain the data's integrity in the metaverse.

The metaverse is also susceptible to cyber-attacks such as hacking and identity theft \cite{di2021Metaverse}, as illustrated in Figure \ref{Fig:MetaverseAndMLthreats}. To mitigate these threats, metaverse platforms should incorporate strong privacy measures like firewalls, intrusion detection systems, and secure communication protocols. Additionally, transparency regarding data collection and usage practices and a commitment to prioritizing user privacy are essential in establishing a safe and secure metaverse environment for all individuals.

\subsection{Homomorphic Encryption (HE)}

Enhancing privacy in AI-XR metaverse is a complex task involving technical and non-technical solutions. From a technical perspective, one approach to improving privacy in AI-XR metaverse is to use secure computation techniques to process sensitive data privately. One mathematical approach for secure computation is the use of HE. HE allows computations to be performed on encrypted data without requiring the data to be decrypted. The computation is performed on ciphertext, and the result is returned as ciphertext, preserving the privacy of the data. Using HE in the metaverse involves encrypting sensitive data before sending it to a server for processing. The server can then perform computations on the encrypted data and return the result as encrypted data, preserving the privacy of the data. The use of HE in AI-XR metaverse can be represented mathematically as follows:

\begin{equation}
 C_i = Enc(x_i) 
\end{equation}
where $Enc$ is the encryption function and $C_i$ is the encrypted data.
\begin{equation}
 y = F(C_1, C_2, ..., C_n) 
\end{equation}
where $F$ is the function to be computed and $y$ is the result.
\begin{equation}
 Dec(y) = F(x_1, x_2, ..., x_n) 
\end{equation}
where $Dec$ is the decryption function.

\begin{figure}[!h]
\centering
\includegraphics[width=0.48\textwidth]{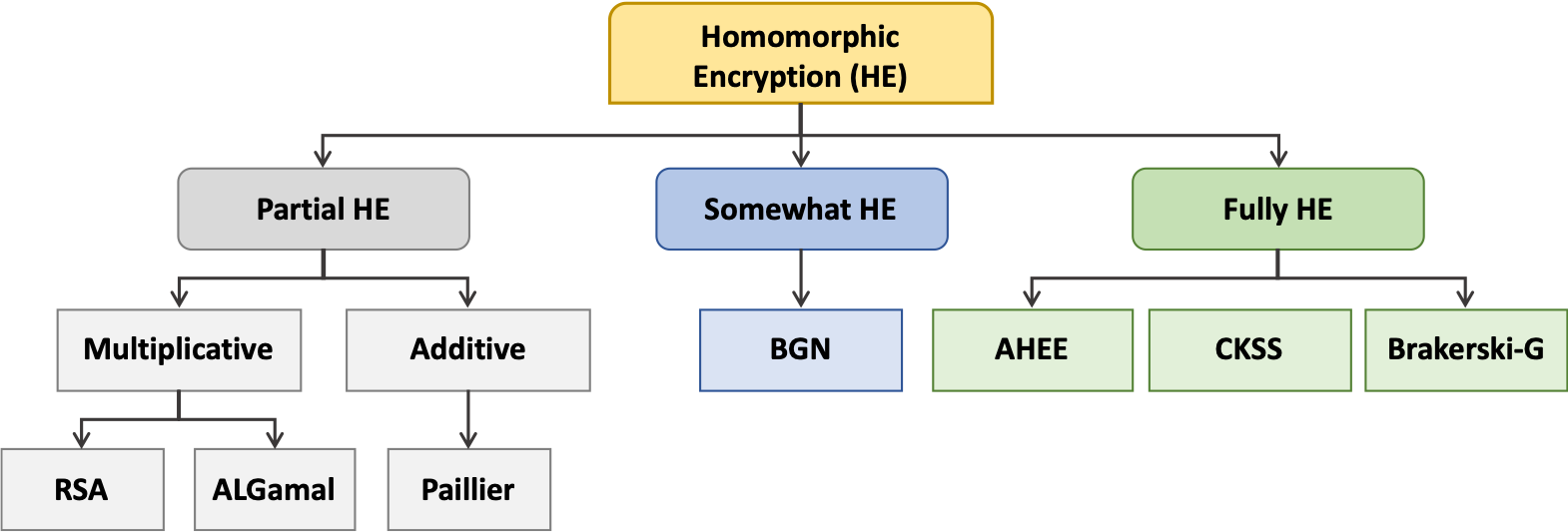}
\caption{Types of homomorphic encryption enabling specific operations on encrypted data without revealing the underlying plaintext.}
\label{fig:HEtypes}
\end{figure}

This mathematical representation shows how computations can be performed on encrypted data, preserving the privacy of the inputs. By using HE, sensitive data can be protected in AI-XR metaverse, enhancing privacy for users. This is just one aspect of improving privacy in AI-XR metaverse, and many other technical and non-technical solutions should also be considered. In HE, two different keys are involved in the encryption and decryption process, a public key and a private key. The public key is made available to anyone who wants to encrypt a message, and the private key is kept secret and only known to the owner. However, the public key is not used for encryption only. Public key cryptography allows mathematical operations to be performed on encrypted data, also known as homomorphic operations. The result of the homomorphic operations remains encrypted, and only the possessor of the private key can decrypt it. Thus, having access to the public key does not allow an attacker to extract the private key, nor does it allow them to read the original data \cite{rivest1978method}. However, it does enable them to perform homomorphic operations on the encrypted data, which may leak some information about the original data. We provide a visualization of the categorization of HE schemes proposed by Armknecht et al. \cite{armknecht2015guide} in Figure \ref{fig:HEtypes}.  

HE has been described as the holy grail of encryption because it can solve many longstanding problems in cryptography. In traditional encryption, data is encrypted before being sent to a third party for computation, but it requires that the data should be decrypted before performing any computation. This means that the data is vulnerable to attacks while it is being processed. In contrast, HE enables computations to be executed directly on encrypted data, eliminating the decryption requirement, i.e., the data remains encrypted and protected throughout the computation process, providing a higher level of security and privacy \cite{tourky2016homomorphic}.

HE has the potential to offer privacy for both the client and server sides, which can be beneficial for many different types of applications. With HE, a client's data can be encrypted before being processed by the server, i.e., the client data is protected and remains private throughout the process. Furthermore, many of the operations involved in NN and ML are HE-friendly, meaning that they can be performed efficiently using HE techniques. These operations include common arithmetic operations such as addition and multiplication, which are essential for training and inference in neural networks.

\begin{figure}[!t]
\centering
\includegraphics[width=8cm]{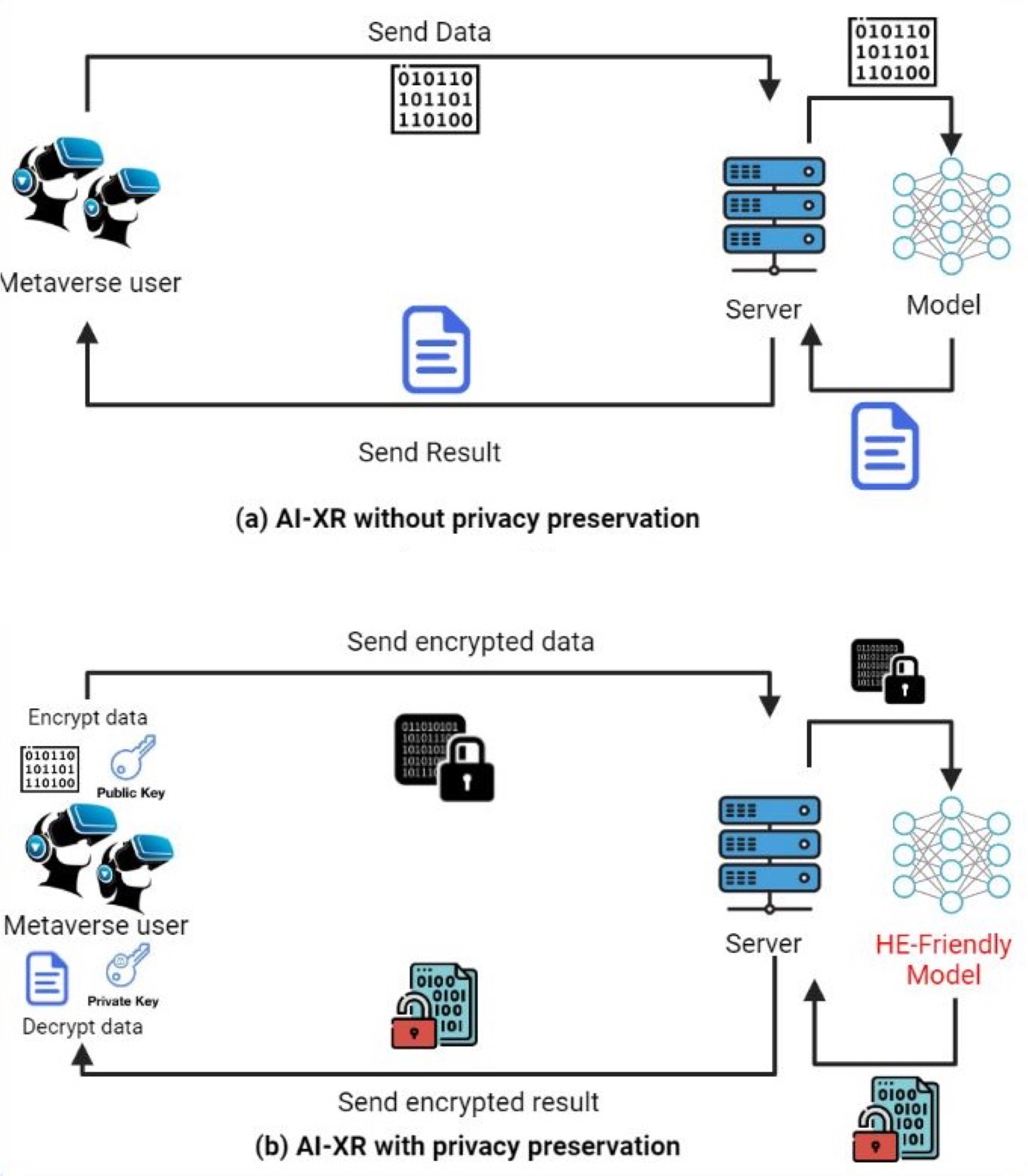}
\caption{Metaverse with and without privacy-preserving using Homomorphic Encryption (HE)}
\label{fig:HEinMetaverse}
\end{figure}

When a client utilizes the resources provided by a service provider (server), a common trust issue arises where the client must rely on the server to handle their data appropriately and solely in the agreed-upon manner. Without privacy preservation techniques, the client must transmit their data in plain form to the server, which is then subjected to processing by the ML model. It creates a potential risk that the server may mishandle or misuse the data in some way, either intentionally or unintentionally, as shown in Figure \ref{fig:HEinMetaverse} (a).

One way to safeguard client data privacy is through the use of HE. With HE, the client encrypts their data before sending it to the server for processing, ensuring it remains under the client's control and is never transmitted in plaintext. HE can protect clients' data by encrypting it before sending it to the server. In this case, the server only receives encrypted data and performs HE computations without accessing the plaintext data. The result of the computation is also encrypted, ensuring complete data privacy. Additionally, the client can use the server to perform encrypted training of a model by transferring an encrypted dataset, allowing the server to train a neural network using encrypted data. However, this approach is relatively rare due to the high computational complexity of training an encryption-compatible neural network, particularly on encrypted data (Figure \ref{fig:HEinMetaverse} b).

The use of HE can address privacy concerns, but it also presents computational challenges. Therefore, finding a balance between privacy, security, and usability is vital during the development of metaverse applications. Al-Ghaili et al. \cite{al2022review} highlight the risk of sensitive information exposure to malicious actors when shared among users. Various privacy-enhancing techniques have been proposed to mitigate this risk, such as the haze of copies technique. This approach aims to confuse adversaries by creating multiple copies of an avatar, making it difficult for them to monitor and gather sensitive information. However, relying solely on this technique may not suffice to completely protect privacy in the metaverse. Additional measures may be necessary, including encryption, access control, and user education.

Sarkar et al. \cite{sarkar2023privacy} presented that HE has the potential to revolutionize the field of privacy-preserving computation by enabling the processing of encrypted data without first having to decrypt it. This can help protect sensitive information from theft or unauthorized access. Currently, research is ongoing to develop more efficient HE algorithms and implementations to reduce computational overheads and make HE more accessible to a wider range of applications. 

\subsubsection{Partially Homomorphic Encryption (PHE)}
This scheme refers to the most basic HE schemes that only provide support for a restricted set of circuits. This is because they can perform computations on encrypted data that involve only addition or multiplication, but not both. PHE is a type of encryption that allows for computations on ciphertext that preserve some of the underlying structure of the plaintext but not all of it. Specifically, a partial HE scheme supports one type of mathematical operation, either addition or multiplication, on encrypted data while preserving the underlying structure of the plaintext. For example, in a PHE scheme that supports addition, given two ciphertexts that respectively encrypt two plaintexts $x$ and $y$, one can perform an addition operation on the ciphertexts to get a new ciphertext that encrypts the sum of $x$ and $y$. However, if we try to perform a multiplication operation on the ciphertexts, we will not get a ciphertext that encrypts the product of $x$ and $y$. PHE has many practical applications, especially in scenarios where privacy is a concern and computations need to be performed on encrypted data. However, the limitations of PHE mean that it is not suitable for all types of computations. Fully Homomorphic Encryption (FHE) schemes, which allow both addition and multiplication operations on ciphertexts, offer more versatility but are still an area of active research due to their computational complexity \cite{armknecht2015guide,paillier1999public}.
\\
\subsubsection{Somewhat Homomorphic Encryption (SHE)}
This is a HE Scheme that can perform computations on encrypted data involving both multiplication and addition. Nevertheless, the size of the ciphertexts increases with each computation performed. Furthermore, the depth of the supported circuits can be managed by adjusting the encryption parameters. For example, in a SHE scheme with a depth of 10, one can perform up to 10 computations involving addition and multiplication on ciphertexts while preserving the underlying structure of the plaintext. However, if we try to perform more than ten computations, the noise in the ciphertext becomes too large, making it difficult to decrypt it accurately. SHE has numerous practical applications, especially in scenarios where data privacy is a concern and computations need to be performed on encrypted data. Compared to PHE, SHE is more versatile as it supports addition and multiplication operations on ciphertexts. However, it is still limited in terms of the number of computations that can be performed, making it unsuitable for certain types of computations \cite{armknecht2015guide,brakerski2011fully}.

\subsubsection{Leveled Homomorphic Encryption (LHE)}
These encryption schemes resemble SHE schemes, with the additional demand that the ciphertext size does not grow during the execution of operations. The parameter $d$ is utilized to regulate the level of circuit depth that can be assessed with LHE schemes. At the same time, the size of the ciphertexts must be unrelated to $d$ and only associated with the security level. When LHE schemes support addition and multiplication operations, they are often termed Leveled Fully Homomorphic Encryption schemes. \cite{podschwadt2022survey}.

\subsubsection{Fully Homomorphic Encryption (FHE)}
\begin{figure}[!h]
\centering
\includegraphics[width=8cm]{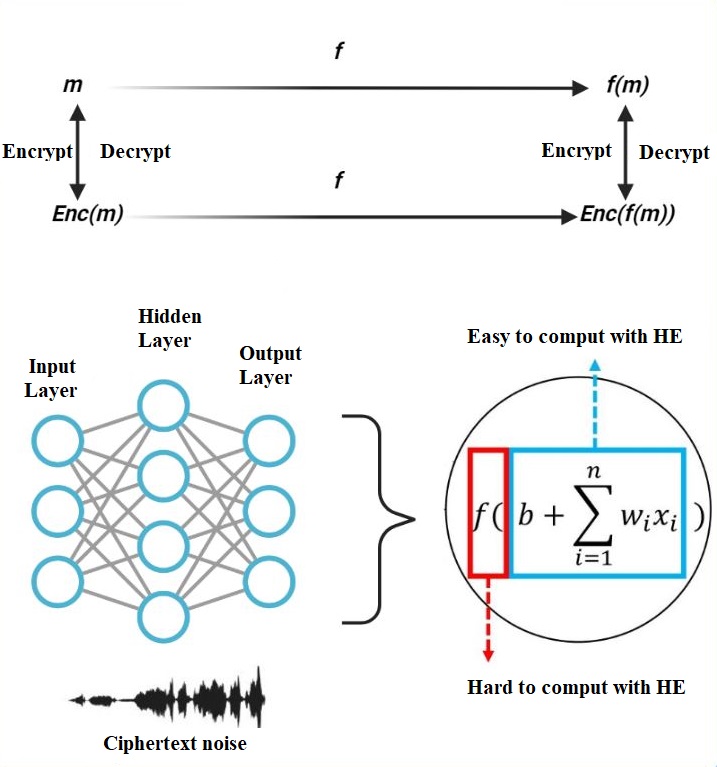}
\caption{Fully Homomorphic Encryption.}
\label{Fig:FullyHE}
\end{figure}

FHE can evaluate circuits without any restrictions on depth or operations (see Figure \ref{Fig:FullyHE}). However, there is a limitation on circuit depth in other schemes due to the encryption process. Encryption adds noise to the plaintext, which makes it harder to decrypt correctly. Each operation executed on encrypted data contributes to the noise within the ciphertext. If the noise surpasses a specific threshold, decryption becomes infeasible. A possible resolution entails decrypting the ciphertext and subsequently re-encrypting it, thereby generating a new ciphertext with a refreshed noise level. Nevertheless, this strategy necessitates possessing the secret key \cite{podschwadt2022survey,han2020better}. FHE has numerous potential applications, such as secure cloud computing, ML on encrypted data, and secure data sharing. 

\subsection{Differential Privacy (DP)}

\begin{figure}[!t]
\centering
\includegraphics[width=8cm]{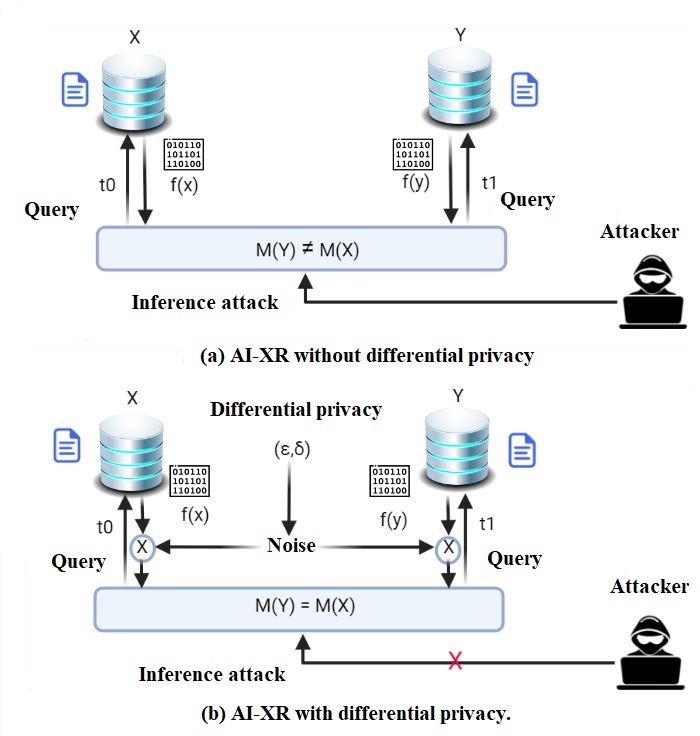}
\caption{AI-XR with and without Differential Privacy.}
\label{Fig:DPinAI-XR}
\end{figure}

Blanco-Justicia et al. \cite{blanco2021achieving} review the use of the DP technique. DP is a privacy-preserving technique used in ML to protect sensitive information in datasets. DP operates by introducing noise to the dataset, which obstructs an attacker's ability to ascertain the contribution of individual data points to the overall results. It preserves individual data points' privacy while still facilitating the analysis of extensive datasets. The DP guarantee is achieved through mathematical proofs showing that privacy breach risk is kept below a specified threshold, Figure \ref{Fig:DPinAI-XR}. This makes DP a powerful tool for protecting sensitive information in datasets, regardless of the resources or motivations of an attacker. DP is an essential tool for maintaining privacy in ML, and it has been widely adopted. DP in ML helps to protect sensitive information in training data and ensures that ML models are developed and used responsibly and ethically \cite{9044259}. It works with standard numerical data types, allowing for widespread support by most ML libraries and hardware accelerators. As a result, implementation and run-time overhead are minimal. However, using DP may impact the quality of the model's predictions. It is important to note that DP does not provide any cryptographic assurances and only provides a probability estimate of potential information leakage \cite{abadi2016deep}. However, balancing privacy and accuracy remains a challenge, and trade-offs often need to be made between the level of privacy protection and the utility of the models.

Here is the formal definition of DP: a randomized algorithm $M$ is said to be $\epsilon$ deferentially private if, for all datasets $x$ and $y$ that differ by at most one element, and for all subsets of outputs $S$, the following condition holds:

\begin{equation}
 Pr[M(x) \in S] \leq e^\epsilon Pr[M(y) \in S] + \delta 
\end{equation}

In this context, $S$ denotes the complete range of potential outputs that $ M$ could anticipate. $x$ denotes the entries in the database, while $y$ represents entries in a parallel database. The parameter $e$ describes the maximum distance between a query on the database ($x$) and an equivalent query on the parallel database ($y$). Additionally, $\delta$ denotes the likelihood of accidental information leakage.

Table \ref{tab:privacy-preserving approaches adv & disadv}
compare and contrast different privacy-preserving approaches for the AI-XR-enabled metaverse proposed in the literature. Not that each approach has advantages and disadvantages, as outlined in Table \ref{tab:privacy-preserving approaches adv & disadv}.

\begin{table}[h]
\centering
\caption{Comparison of various privacy-preserving approaches in terms of their pros and cons.}
\label{tab:privacy-preserving approaches adv & disadv}
\scalebox{0.85}{
\begin{tabular}{|p{1.5cm}|p{3.7cm}|p{3.5cm}|} 
\hline
\textbf{Approach}  & \textbf{Advantages} & \textbf{Disadvantages} \\ \hline

Differential Privacy
&
- Provides strong privacy guarantees\newline
- Can be applied to various types of data\newline
- Well-established approach with a long history of research\newline
&
- Require significant computational resources\newline 
- Reduce the accuracy of AI models\newline
- Hard to implement in some cases\newline
\\
\hline
FL
&
- Allows multiple parties to collaborate and contribute data without sharing the data itself\newline 
- Can improve the privacy of AI models by keeping data on users' devices\newline 
- Can lead to better AI models by leveraging more diverse data\newline
&
- May require significant communication resources\newline 
- May be vulnerable to certain types of attacks, such as poisoning attacks\newline 
- May result in slower training times\newline
\\
\hline
HE
&
- Allows secure computation of encrypted data without decryption\newline
- Can be applied to various types of data and AI models\newline
- Provides strong privacy guarantees\newline
&
- Require significant computational resources\newline 
- May reduce the accuracy of AI models\newline
- Difficult to implement in some cases\newline
\\
\hline
Secure Multi-Party Computation
&
- Allows secure computation of data across multiple parties without revealing the data itself\newline 
- Can be applied to various types of data and AI models \newline
- Provides strong privacy guarantees\newline
&
- May require significant computational resources \newline
- Difficult to implement in some cases \newline
- May result in slower computation times\newline
\\
\hline
Blockchain-based Solutions
&
Provides a decentralized and transparent way to manage data
&
Can be computationally expensive and may require high storage overhead
\\
\hline
\end{tabular}}

\label{tab:table_label}
\end{table}

\subsection{Privacy preservation for AI-XR-enabled metaverses}

With the increasing integration of AI and XR technologies in the metaverse, there is a need to implement privacy-preserving measures to protect user data and ensure trust in the system. One way to preserve privacy in the AI-XR-enabled metaverse is by anonymizing personal data such as user location, behavior, and preferences. This process helps protect user identity and prevent data breaches. Another way to protect user privacy in the AI-XR-enabled metaverse is by applying DP techniques to AI algorithms. It helps to introduce random noise into the data, which in turn helps to prevent individual identification and protect user data. FL is another effective method to preserve privacy in AI-XR-enabled metaverses. By keeping the data decentralized instead of centralizing it in one location, FL reduces the risk of data breaches and protects user privacy. Encryption is also an essential process for preserving privacy in the AI-XR-enabled metaverse. It can be used to protect user data by converting it into an unreadable format, which can only be decrypted using a decryption key. It helps in ensuring that user data remains confidential and secure. Besides, User consent and control are other important factors in preserving privacy in the AI-XR-enabled metaverse. User consent mechanisms, such as privacy policies and terms of service agreements, should clearly outline how user data will be used and protected. Some methods to preserve privacy in an AI-XR-enabled metaverse are shown in Table \ref{tab:ML-ProtectionsApproaches}, which summarizes different approaches that can be leveraged to protect ML models from attacks.

\begin{table*}[]
\centering
\caption{{Overview of methodologies for safeguarding the privacy of ML models against potential attacks.}}
\label{tab:ML-ProtectionsApproaches}
\scalebox{0.76}{
\begin{tabular}{|p {0.5cm}|p {1.7cm}|p {4.8cm}|p {2.5cm}|p {2.5cm}| p {4.3cm} |p {4cm} |}

\hline
\textbf{Ref} & \textbf{Approach} & \textbf{Functionalities} 
& \textbf{Defend Against } & \textbf{Attack Surface } &\textbf{Descriptions - Remarks}
& \textbf{Limitations} \\ \hline
 \cite{10.1145/3436755}
&
Adversarial training 
&
By enhancing the model's resilience, increases the difficulty for attackers to discover inputs capable of triggering misclassifications. 
&
Data evasion and poisoning 
&
Both data and model parameters
&
Challenging to counteract label flipping attack.
&
Challenging to counteract label flipping attack.
 \\ \hline
  \cite{BRAIEK2020110542}
&
Input-transformations.
&
It enhances input data in various ways to improve the privacy of ML models.
&
Adversarial attacks, including evasion and poisoning
&
Input data
&
Involve preprocessing the input to a model.
&
Sensitivity to parameter selection,
Data leakage
 \\ \hline
\cite{10017768}
&
Data-sanitization techniques.
&
Data masking, data perturbation, and data generalization.
&
Membership inference, attribute inference, and more
&
Training data
&
It removes or modifies potentially private data from the training data. This can include techniques such as data masking, data perturbation, and data generalization. 
&
Complexity, difficulty in evaluating model fairness
 \\ \hline
\cite{10.1145/3470496.3527392}
&
Secure Multi-Party Computation (MPC).
&
MPC empowers multiple parties to collaboratively compute a function over their individual private inputs without disclosing them to one another to protect  data from being accessed by other parties during the training process.
&
Data leakage, inference attacks, and collusion 
&
Data, models
&
MPC is used to protect sensitive data from being accessed by the model or other parties during the training process.
MPC ensures that sensitive data is protected even when it is used for training ML models.
&
Computational overhead, communication, and latency
\\ \hline
 \cite{WANG2022446}
&
DP techniques.
&
Adds noise to the data to prevent individuals from being re-identified. 
&
Membership inference, model extraction, and more.
&
Data, predictions, and model training process.
&
A mathematical framework that enables the assessment of the level of privacy afforded by a given system.
&
Complexity and expertise.
\\ \hline
\cite{jin2023fedml}
&
HE
&
Performing computations on encrypted data while maintaining privacy.
&
Data inference attacks.
&
Data, models.
&
HE allows computations on encrypted data without decrypting it, preserving privacy.
&
Slower computation due to encryption and decryption operations, limited support for complex operations and large-scale models.
\\ \hline
\cite{rasha2023federated}
&
Federated Learning (FL)
&
Distributed learning-Training ML models on decentralized devices, keeping data on the local device
&
Data exposure attacks
&
Data, models.
&
FL aggregates model updates from multiple devices while preserving data privacy.
&
Communication overhead, potential data bias, and requires reliable network connections.
\\ \hline
\cite{lee2022privacy}
&
Model Stacking
&
Combining multiple models to make predictions and mitigate single-model attacks
&
Model poisoning attacks
&
Models.
&
Reduces the impact of malicious inputs and improves robustness.
&
High computational and storage requirements and potential overfitting.
\\ \hline

\end{tabular}}
\end{table*}

\subsubsection{Differential Privacy (DP), Memory of ML Models}
DP is a technique used in ML to ensure that individual data points are protected while still allowing for analysis of the data as a whole. It accomplishes this by adding noise to the data before it is analyzed. This noise is carefully controlled so that it does not significantly alter the data analysis, but it makes it much more difficult to identify individual data points. One application of DP is in the memory of ML models. This memory can contain information about the dataset that the model was trained on, which could potentially be used to identify individual data points. Using DP to add noise to the model, the memory can be protected, and individual data points cannot be easily identified. DP is becoming increasingly crucial in ML as more data is collected and analyzed. As data privacy concerns grow, ML models must be trained in a way that protects the privacy of individual data points. DP is a powerful technique for achieving this goal \cite{zhao2019differential}.

Notwithstanding, there remains a potential danger of the model inadvertently disclosing sensitive information during training. Also,  malicious third parties can potentially exploit vulnerabilities in the model to extract the underlying training data. To mitigate this risk, techniques like DP are employed to guarantee that the model cannot disclose any sensitive information. This example illustrates how FL models could potentially leak sensitive information if the FL model for the keyboard's auto-completion feature has unintentionally learned the pattern of credit card numbers from the training data and is then using this information to suggest credit card numbers based on the input text. The model may have learned this information even if the credit card numbers were not explicitly part of the training data. 

To avoid this risk, Generative Adversarial Networks (GANs) can be utilized. These models generate realistic-looking data, such as images or text, based on patterns learned from a training dataset. GANs use two neural networks, a generator network, and a discriminator network, to generate data that closely resembles the training data. The generator network generates data, while the discriminator network distinguishes between real and generated data. Both networks are trained in an adversarial manner where the generator network attempts to deceive the discriminator network, while the discriminator network aims to identify real data accurately. GANs have many applications, such as image and video synthesis, data augmentation, and anomaly detection. However, the potential for misuse of GANs to create fraudulent data and bypass security systems must also be considered \cite{hitaj2017deep}.

\subsubsection{Secure Aggregation}
While DP can prevent the reconstruction of individual training samples, it is still possible to make general conclusions about the underlying data by analyzing the model parameters \cite{fereidooni2021safelearn}. SMPC can be used to ensure that the curator can aggregate the model parameters without knowing the contribution of individual participants. This helps to protect the privacy of the participants and their data. SMPC is a cryptographic method that securely computes data distributed among multiple parties. It ensures the confidentiality of the participants and their data by allowing them to compute a function on their private inputs without revealing them to each other or any third party. The inputs are encrypted in such a way that the computation can be performed without the need for decryption. SMPC is used in various applications, including FL, to ensure that the privacy of the participant's data is maintained while still enabling the computation of a shared result \cite{zhao2019secure}.

\subsubsection{Federated Learning}
FL can potentially safeguard users' privacy in the AI-XR extended metaverse while improving ML models' performance and quality. It is a well-established technology that has gained significant attention recently due to its ability to tackle privacy and scalability challenges in ML \cite{liang2020think}. FL is a collaborative ML technique that enables multiple parties to train a model without sharing their data \cite{qayyum2022collaborative}. Instead, each party trains the model on their local data and transmits only the model updates to a central server. The server then performs parameter aggregation and shares the updated weights with all participating parties. This iterative process continues until the model reaches a satisfactory level of accuracy \cite{li2019fair,du2018efficient}. 

However, protecting data privacy and security is a crucial aspect of FL, as sensitive information may be shared during the model training process, and technical improvements are needed to ensure data protection and prevent data exposure \cite{du2018efficient,yang1038,10.1007/978-3-030-90019-9_11}. FL approach can help protect the users' privacy in the metaverse by minimizing the risk of data breaches, unauthorized access, or misuse of personal information. For example, in virtual or AR environments, users may generate a large amount of sensitive data that can be used to create personalized experiences \cite{aledhari2020federated}.

Data heterogeneity refers to using diverse data sources for ML model training, with communication overhead as the cost of transmitting updates between the central server and remote devices, posing a challenge \cite{li2020preserving}. Connectivity costs also challenge the central server's communication with remote devices or nodes \cite{du2018efficient}, necessitating reduced data transmission to minimize overhead. This may affect model accuracy if data updates are infrequent. Federated Learning (FL) for AI-XR keeps data locally on devices to protect privacy, but privacy challenges remain \cite{seif2020wireless,liu2019boosting}. Training data with missing values for some features can be problematic for ML algorithms requiring complete data \cite{yao2019towards}.

Federated Averaging (FedAvg) is a popular algorithm for distributed training in FL designed to handle large-scale datasets that are spread across many clients. FedAvg can handle large-scale datasets that are distributed across a large number of clients while maintaining privacy \cite{sun2022decentralized}. Since the clients keep their data locally, they do not need to share their data with the central server, which helps to protect the privacy of the clients. The Federated Averaging algorithm is a commonly used mathematical FL model, where local models are aggregated by weighting each one based on the number of available training samples. This accounts for the unequal distribution of data among participants \cite{mcmahan2017communication}. As depicted in the following equation, participants with more data significantly impact the overall model while still considering the input of participants with fewer data. This ensures a fair balance of influence among all participants, regardless of their data quantity.

\begin{equation}
 \theta_{t+1} = \sum_{i=1}^K (n_k/n) * \theta_{t+1}^k 
\end{equation}

where, $\theta_{t+1}$ is the updated global model after $t$ communication rounds,
$\theta_i$ is the local model at device $i$ at the end of round $t$, while 
$n_k$ represents the samples of participants $k$, and  
$n$ is the samples of all participants, $\theta_{t+1}^k$ is the local model parameter of participants $k$.

\subsection{Solutions for AI Associated Privacy Issues}

\subsubsection{Privacy and Security in ML Models for Data Sharing}

Weng et al. \cite{zhang2021adaptive} used a convolutional neural network CNN-based blockchain framework to ensure data privacy and integrity in a network. Blockchains are secure and tamper-proof, making them well-suited for applications where privacy and data integrity are essential. Additionally, Convolutional Neural Networks (CNNs) can process and analyze data in a blockchain network, offering additional security and privacy protection. Combining these two technologies can result in a highly secure and private data sharing and management system in a decentralized network.

\subsubsection{Collaborative ML Approaches for Privacy}

Soykan et al. \cite{soykan2022survey} presented that collaborative ML approaches like FL allow organizations to train models on data distributed across different devices or entities without requiring the data to be centralized or aggregated. This helps to address privacy and security concerns, as the sensitive data remains on the device and is not disclosed to the central server or any other entity. The model parameters are instead communicated and averaged between the devices, allowing for a collaborative training process while preserving the privacy of the data.

\subsubsection{Privacy-Preserving Face Recognition System}

Wang et al. \cite{wang2019privacy} proposed a privacy-preserving face recognition system based on edge computing and nearest-neighbor encryption. The system aims to balance privacy and security by extracting facial features using a CNN model and encrypting the feature vectors to preserve privacy. The cosine similarity is then calculated over the encrypted vectors to identify persons, with edge computing utilized to transfer some operations from the cloud to the edge nodes for increased efficiency.

\subsubsection{Privacy-Preserving Outsourced SVM Classification}

Liang et al. \cite{liang2019efficient} designed a system for privacy-preserving outsourced SVM classification developed using order-preserve encryption. The system focuses on protecting the privacy of the requester's data and the provider's classifier by utilizing encryption. They employed order-preserve encryption for encrypting the data to enable meaningful computation and analysis of the encrypted data while maintaining privacy. This approach facilitates privacy-preserving outsourced SVM classification while preserving the order of the data.

\subsubsection{Distributed Transfer Learning for Privacy in Activity Recognition}

Hashemian et al. \cite{hashemian2019privacy} introduced a distributed transfer learning scheme for activity recognition applications that preserves privacy. The method adapts the training mode based on the trust mode of the participant client and server. The system consists of two privacy-preserving transfer learning algorithms, one designed to preserve data privacy on the client side and the other to protect privacy at both the client and server levels. The system aims to facilitate the transfer of knowledge from one model to another while safeguarding the training data's privacy.

\begin{table*}[!t]
\centering
\caption{Overview of proposed AI-based solutions for addressing privacy issues.}
\label{tab: Summarize all AI Associated proposed solutions Privacy problems}
\scalebox{0.9}{
\begin{tabular}{|l|p {7.5cm}|p {4.78cm}| p {4cm} |}
\hline
\textbf{Authors}  & \textbf{Method-Description} & \textbf{Application} & \textbf{Limitation(s)} \\ \hline
Weng et al. \cite{zhang2021adaptive}
&
CNN-based blockchain framework - providing a decentralized and secure infrastructure for data sharing and model training
&
Data privacy and integrity in a decentralized network
&
Network latency
\\
\hline
Soykan et al. \cite{soykan2022survey}
&
FL- striking a balance between collaborative model training and individual data privacy.
&
Training models on distributed data while preserving the privacy
&
Data Fragmentation, 
Communication Overhead
\\
\hline
Wang et al. \cite{wang2019privacy}
&
Privacy-preserving face recognition - anonymization and pseudonymization.
&
Balancing privacy and security in face recognition using edge computing
&
Data Storage,
Network bandwidth, Limited  processing power
\\
\hline
Liang et al. \cite{liang2019efficient}
&
Outsourced SVM--allowing entities to train SVM for classification tasks while protecting the privacy of their sensitive data.
&
Protecting data privacy and preserving SVM classification using encryption
&
Data leakage
\\
\hline
Hashemian et al. \cite{hashemian2019privacy}
&
Distributed transfer learning allows entities to collaborate on model training while preserving the privacy of their own data.
&
Privacy-preserving activity recognition
&
Loss of data granularity, Heterogeneity of data
\\
\hline
Huang et al. \cite{huang2011adversarial}
&
Robust and secure algorithms -by incorporating privacy-preserving techniques and enhancing the security of AI systems.
&
Resilient to data poisoning attacks in ML
&
Computational overhead, Data dependency, and distribution
\\
\hline
Sun et al. \cite{sun2021data}
&
Defense against gradient ascent attack - by protecting the privacy and confidentiality of sensitive information stored in AI models.
&
Resilient defenses against attacks in SVM and graph-based models
&
Limited robustness, Increased computational complexity 
\\
\hline
Hussain et al. \cite{hussain2020machine}
&
ML and DL techniques -  incorporate data anonymization methods to remove or obfuscate personally identifiable information (PII) from datasets, reducing the risk of data re-identification.
&
Security challenges in IoT networks
&
Computational overhead
\\
\hline
Guo et al. \cite{guo2021federated}
&
FL - allows healthcare organizations to keep their data locally stored within their own infrastructure
&
Healthcare applications, such as predictive models and personalized treatment plans
&
Data leakage, Differential privacy trade-offs, Model inversion attacks.
\\
\hline
Alazab et al. \cite{alazab2021federated}
&
FL-by enabling privacy-preserving model training and inference directly on the edge devices themselves
&
Edge devices, 
&
Data leakage, Differential privacy trade-offs, Model inversion attacks.
\\
\hline
Mohammed et al. \cite{aledhari2020federated}
&
FL -allows other organizations to keep their data locally stored within their own infrastructure
&
Healthcare or financial data and AI-XR metaverse applications.
&
Data leakage, Differential privacy trade-offs, Model inversion attacks
\\ \hline
\end{tabular}}
\end{table*}

\subsubsection{Security Concerns with Data Poisoning Attacks}

Huang et al. \cite{huang2011adversarial} presented that data poisoning attacks have become a major concern in ML. In these types of attacks, an attacker manipulates a small portion of the training data to cause the ML model to produce incorrect or harmful predictions. As ML models are increasingly being used in security-sensitive applications, it has become imperative to develop such secure and robust algorithms that are resilient to these attacks. 

\subsubsection{Gradient Ascent Attack and Data Poisoning}

Sun et al. \cite{sun2021data} highlighted a concerning type of attack called the gradient ascent attack against SVM models, which involves computing a gradient based on the SVM's optimal solution and modifying the model's parameters using a gradient ascent technique. The objective is to cause the SVM to make erroneous or harmful predictions, posing serious risks to the security and reliability of ML models, necessitating robust defenses against such attacks. Additionally, the authors discussed data poisoning attacks and their extensive study on various ML models, including matrix factorization-based collaborative filtering for autoregressive models and neural networks for graph data. Autoregressive models \cite{alfeld2016data}, used in statistics and econometrics, model time-varying processes where a variable's current value depends on its past values. These models are represented as linear equations, estimating coefficients through methods like maximum likelihood estimation. AR models find widespread applications in economics \cite{jiao2019tourism}, finance \cite{xu2021assessing}, and engineering \cite{delgado2020research}, used for forecasting, signal processing, and pattern recognition \cite{ouali2022augmented}.

\subsubsection{ML and DL Techniques for IoT Security}

Hussain et al. \cite{hussain2020machine} presented the potential of ML and DL techniques in providing embedded intelligence for IoT devices and networks. They can be employed to tackle a range of security challenges, such as detecting anomalies and intrusions in network traffic, identifying malicious activity in IoT devices, and performing threat analysis. DL techniques can also be utilized for image and speech recognition, sentiment analysis, and NLP, which can enhance the security of IoT systems. Furthermore, ML and DL algorithms can learn behavioral patterns and make predictions that can facilitate the development of proactive security strategies and improve decision-making in IoT security. Nevertheless, it is crucial to recognize that ML and DL algorithms may be vulnerable to attacks, requiring appropriate measures to ensure IoT system security.

\subsubsection{Privacy Concerns and Solutions in FL}

Privacy is a major concern in FL, as highlighted by Aledhari et al. \cite{aledhari2020federated}. The decentralized nature of FL introduces vulnerabilities like data poisoning and model poisoning attacks, enabling adversaries to introduce biases and vulnerabilities \cite{qayyum2022making}. Protecting user privacy becomes critical, particularly when handling sensitive data like healthcare or financial information. FL holds promise for AI-XR metaverse applications, but it requires addressing technical complexities, including limited computational power, communication costs, and privacy and security issues. Efficient communication protocols, robust security mechanisms, and optimized model aggregation techniques can mitigate these challenges. The work by Alazab et al. \cite{alazab2021federated} underscores distributed data privacy as a significant concern in FL, where poisoning and reconstruction attacks can occur. Data poisoning manipulates training data to introduce biases, while reconstruction attacks exploit updates sent from edge devices to the central server, resulting in biased models and affecting system performance. Robust security mechanisms are vital to prevent such attacks and preserve data privacy. Integrating privacy-preserving techniques within FL can further enhance resilience against privacy attacks, such as leveraging HE for tasks like detecting textual-based misinformation, as demonstrated by Ali et al. \cite{ali2022spam}. To encapsulate all AI-related proposed solutions to privacy issues, consult Table \ref{tab: Summarize all AI Associated proposed solutions Privacy problems}, which summarizes these solutions comprehensively. For more in-depth information, kindly refer to Table\ref{tab:SurveyedPapers}, offering an extensive overview of the Privacy Attacks and Defenses identified within the surveyed papers.

\begin{table*}[!t]
  \begingroup
 \renewcommand*{\arraystretch}{1.5}%
 \definecolor{tabred}{RGB}{230,36,0}%
 \definecolor{tabgreen}{RGB}{0,116,21}%
 \definecolor{taborange}{RGB}{255,124,0}%
 \definecolor{tabbrown}{RGB}{171,70,0}%
 \definecolor{tabyellow}{RGB}{255,253,169}%
 \newcommand*{\headformat}[1]{{\small#1}}%
 \newcommand*{\vcorr}{%
   \vadjust{\vspace{-\dp\csname @arstrutbox\endcsname}}%
   \global\let\vcorr\relax
}%
 \newcommand*{\HeadAux}[1]{%
   \multicolumn{1}{@{}r@{}}{%
  \vcorr
  \sbox0{\headformat{\strut #1}}%
  \sbox2{\headformat{Complex Data Movement}}%
  \sbox4{\kern\tabcolsep\xmark\kern\tabcolsep}%
  \sbox6{\rotatebox{45}{\rule{0pt}{\dimexpr\ht0+\dp0\relax}}}%
  \sbox0{\raisebox{.5\dimexpr\dp0-\ht0\relax}[0pt][0pt]{\unhcopy0}}%
  \kern.75\wd4 %
  \rlap{%
 \raisebox{.25\wd4}{\rotatebox{45}{\unhcopy0}}%
 }%
  \kern.25\wd4 %
  \ifx\HeadLine Y%
 \dimen0=\dimexpr\wd2+.5\wd4\relax
 \rlap{\rotatebox{45}{\hbox{\vrule width\dimen0 height .4pt}}}%
  \fi
}%
}%
 \newcommand*{\head}[1]{\HeadAux{\global\let\HeadLine=Y#1}}%
 \newcommand*{\headNoLine}[1]{\HeadAux{\global\let\HeadLine=N#1}}%
 \noindent
 \begin{tabular}{%
   >{\bfseries}lc|>{\quad}c
   *{4}{c|}c>{\quad}c
   *{4}{c|}c>{\quad}c
   *{2}{c|}c%
}%
   &
   \head{References} &
   &
   \head{Privacy Attack} &
   \head{Privacy Defense} &
   \head{Defense Mechanisms} &
   \head{VR Devices Adversaries} &
   \headNoLine{Hardware Adversaries} &
   &
   \head{Software Adversaries} &
   \head{ML Adversaries} &
   \head{ML  Protection} &
   \head{User Interaction Protection} &
   \headNoLine{Real-time Protection} &
   &
   \head{Data Protection} &
   \head{Net-Protection} &
   \\
   \rowcolor{tabyellow}%
   \sbox0{S}%
   \rule{0pt}{\dimexpr\ht0 + 2ex\relax}%
 Nair & \cite{nair2022exploring}&& MB,ME,UB
  & \xmark& HF-VR-D,RAC & 
  \faVrCardboard 
  \faEye
  \faMicrophone 
  \faWifi
  \faGamepad
  & \cmark &&\cmark & \xmark& \xmark& \cmark 
  & \xmark&& \cmark & \xmark

   \\\hline
  Winkler et al. &  \cite{winkler2022questsim} & & F/N
   & \xmark& RLF & \faVrCardboard \faGamepad
   & \cmark & & \cmark & \xmark & \xmark
   & \xmark&\xmark& & \xmark&  \xmark
   \\\hline
   Keshk et al. & \cite{keshk2019privacy} && Data-P, SCADA &  \cmark & Block-C,Auto-E
   &  \faWifi \faSun &  \cmark && \cmark  & \cmark  & \cmark 
   & \cmark & \xmark&& \cmark & \cmark 
   \\\hline
   Yi et al. &  \cite{yi2022stackelberg} && Data Leakage &  \cmark
   & FL, DP & \faIcon{mobile-alt} \faWifi & \cmark
   && \cmark & \cmark & \cmark & \xmark& \xmark&&
   \cmark & \cmark
   \\\hline
  Qayyum et al.&  \cite{qayyum2022making} && M-Par, Lab-FA&  \cmark
   & FL & \faIcon{mobile-alt} & \cmark
   && \cmark & \cmark & \cmark & \xmark& \xmark&&
   \xmark& \xmark

   \\\hline
  Yang et al.&  \cite{yang2023model} && Data-P, MSA & \cmark
   & FL, DP & \faIcon{mobile-alt} \faWifi & \xmark
   && \cmark & \cmark & \cmark & \xmark& \xmark&&
   \cmark & \xmark

   \\\hline
  Munilla et al.&  \cite{munilla2023sok} && VR-P-Threats &  \cmark
   &Meta-G,VPN,Enc& VR-sensors & \cmark
   && \cmark & \cmark & \cmark & \cmark & \xmark&&
   \cmark & \cmark

   \\\hline
  Zhou et al.&  \cite{zhou2022ppmlac} && S-Ch-A, MITM &  \cmark
   & MPC,FL,HE & \faWifi & \cmark
   && \cmark & \cmark & \cmark & \xmark& \xmark&&
   \cmark & \cmark 

   \\\hline
  Mishra et al.&  \cite{yi2022stackelberg} && I-Th,fraud&  \cmark
   & Block-C& \faWifi & \xmark
   && \cmark & \xmark& \xmark& \xmark& \cmark &&
   \cmark & \cmark 

 \\\hline
  Weng et al.&  \cite{zhang2021adaptive} && Data Tamper &  \cmark
  & CNN-based-B &  \faWifi & \cmark
  && \cmark & \cmark & \cmark & \xmark
  & \xmark&&
 \cmark & \cmark 

   \\\hline
  Wang et al.&  \cite{wang2019privacy} && Data-P &  \cmark
   & NNE, HE & \faIcon{mobile-alt} \faWifi & \cmark
   && \cmark & \cmark & \cmark & \xmark& \xmark&&
   \cmark & \cmark 
 \\\hline
  Mavridou et al.&  \cite{mavridou2018towards} && PPG,ECG-S &  \cmark
   & A-Det-S & \faHeartbeat  \faVideo \faTint& \cmark
   && \cmark & \xmark& \xmark& \xmark& \xmark&&
   \xmark& \xmark

 \\\hline
  Arafat et al.&  \cite{al2021vr} && VR-Spy &  \xmark
   & P-E-Alg & \faWifi & \cmark
   && \xmark& \xmark& \xmark& \xmark& \xmark&&
   \xmark& \xmark

 \\\hline
  David et al.&  \cite{david2021privacy} && Eye Tracking &  \cmark
   & MR-Use-Cases & \faEye & \cmark
   && \cmark & \xmark& \xmark& \xmark& \xmark&&
   \cmark & \cmark 

 \\\hline
  Chaud et al.&  \cite{chaudhary2020privacy} && Iris Iden &  \cmark
   & FL, DP & \faEye & \cmark
   && \cmark & \cmark & \cmark & \xmark& \xmark&&
   \cmark & \cmark 

  \\\hline
  Tricomi et al.&  \cite{tricomi2023you} && Mov-Bio &  \cmark
   & ML & 
  \faVrCardboard 
  \faEye
  \faGamepad
   & \cmark
   && \cmark & \xmark& \xmark& \xmark& \xmark&&
   \xmark& \xmark

  \\\hline
  Liebers et al.&  \cite{liebers2021understanding} && Mov-Bio &  \cmark
   & O-Q-HMD & 
  \faVrCardboard  \faCompass

   & \cmark
   && \cmark & \xmark& \xmark& \cmark & \xmark&&
   \xmark& \xmark

  \\\hline
  Al-Ghaili et al.&  \cite{yi2022stackelberg} && Avatar-Tamper&  \cmark
   & HoCT & \faIcon{mobile-alt} \faWifi & \cmark
   && \cmark & \cmark & \cmark & \xmark& \xmark&&
   \xmark& \cmark 

  \\\hline
  Blanco et al.&  \cite{blanco2021achieving} && Data-P &  \cmark
   & DP & MI & \xmark
   && \xmark& \cmark & \cmark & \cmark & \xmark&&
   \cmark & \xmark

  \\\hline
  Zhou et al.&  \cite{10.1145/3470496.3527392} && Data-P &  \cmark
   & MPC & F/N & \xmark
   && \xmark& \cmark & \cmark & \xmark& \xmark&&
   \cmark & \cmark
   
  \\\hline
  Paul et al.&  \cite{liang2020think} && Data-P &  \cmark
   & FL  & \faIcon{mobile-alt} \faWifi & \xmark
   && \xmark& \cmark & \cmark & \xmark& \xmark&&
   \cmark & \xmark

  \\\hline
  Jiwei et al.&  \cite{9695995} && DoS&  \cmark
   & SVMs & MI & \xmark
   && \cmark & \cmark & \cmark & \xmark& \xmark&&
   \cmark & \cmark 

  \\\hline
  Ximeng et al.&  \cite{7047716} && Data-P &  \cmark
   & NBC & MI & \xmark
   && \cmark & \cmark & \cmark & \xmark& \xmark&&
   \cmark & \cmark 

  \\\hline
  Kainat et al.&  \cite{2022} && Data-P &  \cmark
   & KNN & MI & \xmark
   && \cmark & \cmark & \cmark & \xmark& \xmark&&
   \cmark & \cmark

  \\\hline
  Ling et al.&  \cite{ling2019know} && keystrokes  &  \xmark
   & F/N & \faVrCardboard \faGamepad \faVideo \faVolumeUp & \cmark
   && \cmark & \xmark& \xmark& \xmark& \xmark&&
   \xmark& \xmark

  \\\hline
  Maloney et al.&  \cite{maloney2020anonymity} &&Speech-Rec&  \cmark
   & Voice Modul &  \faVolumeUp & \cmark
   && \cmark & \xmark& \xmark& \xmark& \xmark&&
   \xmark& \xmark
   
  \\\hline
  Hussain et al.&  \cite{hussain2020machine} && Net-Traffic &  \cmark 
  & ML, DL & IoT, MI & \xmark
  && \xmark& \cmark & \cmark & \xmark& \xmark&&
   \cmark & \cmark 
  
  \\\hline
  Xiaohui et al.&  \cite{guo2021federated} && Data-P &  \cmark
   & FL, DP & \faHeartbeat \faWifi \faTint & \cmark
   && \cmark & \cmark & \cmark & \xmark& \xmark&&
   \cmark & \cmark 
  \\\hline
  Cheng et al.&  \cite{cheng2022differentially} && Data Leakage &  \cmark
   & FL, DP & MI & \xmark
   && \cmark & \cmark & \cmark & \xmark& \xmark&&
   \cmark & \cmark 
  \\\hline
  Fung et al.&  \cite{fung2018mitigating} && Model-P &  \cmark
   & FL & \faIcon{mobile-alt} \faWifi & \cmark
   && \cmark & \cmark & \cmark & \xmark& \xmark&&
   \cmark & \cmark

   \\[.5ex]
   \rowcolor{tabyellow}%
   \multicolumn{3}{c}{} &
   \multicolumn{5}{c}{\bfseries Limitations} &&
   \multicolumn{5}{c}{\bfseries Adversaries} &&
   \multicolumn{3}{c}{\bfseries Protections}
   \\
 \end{tabular}%
 \kern19.5mm 
  \endgroup
\caption{Details related to privacy attacks and defenses in surveyed papers.
\label{tab:SurveyedPapers}
\newline \newline \textbf{SENSORS}:
Compass: \faCompass; 
Edge device: \faIcon{mobile-alt}; 
Eye trackers: \faEye; 
Hand-Held Controllers: \faGamepad; 
Head-mounted displays (HMDs):\faVrCardboard; 
Heart rate: \faHeartbeat;
Humidity: \faTint;
Microphone:\faMicrophone; 
Network devices:\faWifi; 
Optical trackers \faVideo;
Sound: \faVolumeUp;
\newline
\textbf{ACRONYMS}:
{\scriptsize \textit{Auto-E}: AutoEncoders; 
\textit{Block-C}: Blockchain;
\textit{CNN-Based-B}: CNN-Based Blockchain; 
\textit{Data-P}: data poisoning; 
\textit{DP}: Differential Privacy; 
\textit{ECG}: Electrocardiography signals; 
\textit{Enc}: Encryption; 
\textit{FL}: Federated Learning;
\textit{F/N} = False negative; 
\textit{F/P} = False positive; 
\textit{HE}: Homomorphic Encryption; 
\textit{HF-VR-D}: High-Fidelity VR Devices; 
\textit{HoCT}: Haze of Copies Technique; 
\textit{I-Th}: Identity Theft;
\textit{KNN}: K-Nearest Neighbor; 
\textit{Lab-FA}: Label Flipping Attack; 
\textit{M-Par}: Malicious Parameters; 
\textit{MB}: Metaverse Biometrics; 
\textit{ME}: Metaverse Environment; 
\textit{Meta-G}: Meta Guard; 
\textit{MITM}: Man-in-the-middle attacks; 
\textit{Mov-Bio}: Movement Biometrics; 
\textit{MPC}: Multi-Party Computations; 
\textit{MR-Use-Cases}: MR Use Cases; 
\textit{MSA}: Model Shuffle Attack; 
\textit{NBC}: Naive Bayes classifier; 
\textit{NNE}: Nearest Neighbor Encryption; 
\textit{O-Q-HMD}: Oculus Quest; 
\textit{P-E-Alg}: Pattern Extraction Algorithm; 
\textit{PPG}: PhotoPlethysmoGraphy; 
\textit{RAC}: Random Alphanumeric Codes; 
\textit{RLF}: Reinforcement Learning Framework; 
\textit{S-CH-A}: Side-Channel Attacks; 
\textit{SCADA}: Supervisory Control and Data Acquisition systems; 
\textit{Speech-Rec}: Speech Recognition.
\textit{SVMs}: Support Vector Machines; 
\textit{UB}: User Behavior;
\textit{UL}: User Location; 
\textit{VPN}: Virtual Private Network; 
\textit{VR-DS}: VR Device Specifications; 
\textit{VR-P-Threats}: VR Privacy Threats.
}
}
\end{table*}

\subsubsection{End-to-End ML Pipeline for Private AI-XR Applications}

Developing secure and private AI-XR applications involves several stages in the ML pipeline. Firstly, data collection and preprocessing include gathering data from trustworthy sources while considering privacy regulations and user consent. To protect individual identities, anonymize or pseudonymize personal data and apply data minimization techniques to collect only necessary data and avoid storing sensitive information. Then, during model training, implement privacy-preserving methods like FL or differential privacy. Model evaluation and testing involve assessing the model's security and privacy risks, including vulnerability to attacks or unintended data leaks. When deploying the model, implement access controls and user authentication mechanisms to restrict system access to authorized users and regularly update and patch the deployed system to address privacy vulnerabilities. It involves conducting privacy impact assessments to identify potential risks and mitigate privacy concerns associated with the AI-XR application. By following this comprehensive approach, AI-XR applications can maintain a high level of privacy and security for their users.

\section{Insights and Pitfalls}

This section highlights the benefits of the survey by uncovering critical insights into privacy risks and challenges in AI-XR-enabled metaverses that may not immediately become apparent. While approaches like FL, HE, and DP show promise for privacy preservation in the metaverse, they also have limitations that must be considered. These limitations include computational and communication overhead in FL, performance impact and limited computation capabilities in HE, trade-offs between privacy and utility, and susceptibility to attacks in DP. Researchers and developers must carefully consider these limitations and trade-offs when implementing privacy-preserving methods in the metaverse to ensure adequate privacy protection while maintaining functionality and usability.

\subsection{Distinctive Privacy Challenges in the Metaverse}

The metaverse, a virtual reality space with unique features, shares some privacy concerns with other IT service platforms. However, it is crucial to adopt a tailored security and privacy approach that specifically addresses the distinct characteristics of the metaverse rather than relying solely on pre-existing security measures. The metaverse presents distinctive privacy challenges stemming from its novelty, intricacy, and multisensory environment. With sophisticated elements like 3D graphics and immersive experiences, privacy violations and security breaches within the metaverse can have severe consequences for users and victims. As a result, privacy violations within the metaverse are likely to lead to amplified technical impact \cite{haber2019digital}. In addition to these challenges, the metaverse introduces significant data privacy concerns that require attention. Biased algorithms trained on non-representative data can perpetuate discrimination and inequality. The lack of clear regulations and standards in the metaverse contributes to user uncertainty and increased privacy risks. Furthermore, the interconnected nature of the metaverse amplifies the impact of data breaches, posing widespread privacy and security vulnerabilities.

\subsection{Copyright Protection and Personal Data Management}

Safeguarding copyright within the metaverse requires a specific strategy to protect intellectual property \cite{dwivedi2022Metaverse}. The metaverse's unique characteristics and complexity present challenges in monitoring and detecting attacks compared to existing platforms \cite{alspach2022fate}. Innovative approaches are needed to ensure platform security, integrity, and user protection. The metaverse encompasses a multi-sensory environment with sophisticated elements like 3D graphics and immersive experiences. Managing personal data within the metaverse can benefit from blockchain and smart contracts, offering a secure and transparent approach. Continuous research and development efforts are necessary to stay ahead of data privacy innovations and adapt to the evolving metaverse landscape, including investing in emerging technologies such as new encryption methods (HE and DP) and secure data-sharing protocols. Building strong partnerships with organizations in the metaverse ecosystem is essential for maintaining a safe and responsible virtual environment. Collaboration and cooperation in addressing data privacy challenges will contribute to users' overall well-being and protection in the metaverse.

\subsection{Blockchain and Smart Contracts for Data Management}
Blockchain technology and smart contracts have emerged as promising solutions for enhancing privacy and security in the metaverse \cite{wang2021blockchain, maksymyuk2022blockchain, mishra2022contribution}. By leveraging these technologies, the management of personal data within virtual reality environments can be fortified with robust data protection mechanisms and decentralized control over personal information. It empowers users with increased privacy and security, offering them greater confidence in sharing and interacting within the metaverse. The transparent nature of blockchain and the automated execution of smart contracts provide a trustless environment, reducing the reliance on centralized authorities and minimizing the risk of data breaches or unauthorized access. Incorporating blockchain and smart contracts into the metaverse's privacy framework is a crucial step towards fostering a secure and privacy-preserving virtual reality experience for users.

\subsection{Regulatory Challenges and Policy Considerations}

Governments and regulatory bodies face challenges in addressing data privacy in the evolving Metaverse. Virtual environments' dynamic and cross-border nature requires collaboration to develop consistent regulations and standards. Comprehensive rules and guidelines are needed to govern the collection, storage, and use of personal data in the Metaverse. Balancing innovation and privacy rights is crucial to create a safe and equitable virtual environment.

\subsection{Empowering Users and Ensuring Transparency}

Empowering users with control over their personal information and raising awareness of the risks associated with data sharing in the Metaverse is essential. Algorithmic transparency and accountability guidelines can mitigate the risk of biased outcomes. Proactive measures are needed to protect user privacy and promote a fair and inclusive virtual experience.

\section{Open Research Issues}

\subsection{Human Computer Interaction Issues}

The technology used for interaction in the metaverse links the virtual and real worlds must fulfill certain conditions. These conditions consist of the interactive devices being lightweight, portable, wearable, and transparent so that users can fully immerse themselves in the virtual world without interruptions. Interactive technologies frequently utilized in the metaverse comprise somatosensory technologies (XR, VR, AR, MR), and brain-computer interfaces. Among these, XR technology amalgamates real and virtual environments and includes immersive technologies like VR, AR, and MR. On the other hand, Somatosensory technology enables users to interact with their surroundings and devices through body movements. Nonetheless, existing interactive devices employing these technologies face certain limitations. They are typically not sufficiently lightweight or transparent, and their high cost makes them challenging to adopt widely. Brain-computer interfaces are available in invasive, semi-invasive, and non-invasive forms. Invasive methods involve surgically implanting electrodes into the cerebral cortex, which provides the most accurate results but carries significant surgical risks and risks of tissue rejection. Non-invasive methods use wearable devices attached to the scalp to interpret EEG signals, but the signal collection is relatively weak. Additionally, disseminating brain-computer interfaces presents its own set of challenges \cite{ning2021survey}.

\subsection{Rethinking Privacy in AI-XR Metaverses}
Researchers face important questions in the early stages of researching the privacy of AI-XR-enabled metaverses. One urgent concern is the development of VR games or applications that enable discreet privacy attacks, potentially integrated into everyday tasks in future VR/AR environments. Analytical techniques can be explored to uncover hidden data collection mechanisms and increase the difficulty of executing these attacks. Further investigation into utilizing unexplored data sources, such as eye tracking and full-body tracking, would enhance our understanding of VR-related vulnerabilities. Additionally, studying how attackers could manipulate user opinions rather than solely observing them would shed light on the future implications of immersive metaverse applications. It is crucial to explore countermeasures for these VR privacy attacks, such as introducing noise to raw VR device data while maintaining a seamless user experience. Future work should focus on practical and effective solutions that minimize vulnerabilities while delivering an immersive and engaging VR experience.
\label{sec:open}

\vspace{2mm}
\subsubsection{Consent Management}
Effective consent management is vital for safeguarding privacy in the metaverse, establishing mechanisms to obtain and manage user consent transparently and responsibly. Informed consent requires providing clear and comprehensive information about data types, usage, and potential risks in a user-friendly manner, enabling users to make informed decisions by granting or denying consent for specific data processing activities. Empowering users with control over their data necessitates the ability to modify consent choices easily through user-friendly privacy dashboards or settings, acknowledging that preferences may change over time. Integrating privacy considerations into the design of metaverse platforms and applications is crucial, incorporating privacy-enhancing features and aligning data practices with privacy principles. Educational initiatives are essential for promoting user awareness and understanding of consent and privacy, providing accessible information about the importance of privacy, data implications, and guidance on managing consent settings. Holding organizations accountable for adhering to consent preferences and privacy regulations can be achieved through audits, transparency reports, and regulatory oversight. By prioritizing effective consent management, the metaverse can protect user privacy, foster trust, and empower individuals to control their data.

\vspace{2mm}
\subsubsection{Ethical Use of AI}
In developing and deploying AI algorithms within the metaverse, it is vital to address the ethical implications and uphold principles of fairness, accountability, and transparency. Fairness in ML requires AI algorithms to avoid discrimination based on factors such as race, gender, or socioeconomic status, which can be achieved by mitigating biases through diverse training datasets and fairness metrics. Accountability involves establishing mechanisms to hold developers and organizations responsible for AI algorithm actions, facilitated by transparent governance frameworks and regulatory guidelines. Ensuring recourse and redress mechanisms for individuals affected by AI harm is also essential. Transparency is critical in building trust and understanding AI workings in the metaverse, requiring clear explanations of decision-making processes and data usage. Additional ethical considerations include privacy protection, informed consent, and considering societal impacts. The interdisciplinary collaboration between AI researchers, ethicists, policymakers, and stakeholders is crucial for developing guidelines and best practices that promote ethical AI use in the metaverse. By actively addressing these ethical concerns, we can create a metaverse that respects user rights, fosters trust, and benefits society.

\subsection{Embedded/ Edge ML For Enhanced Data Privacy}

Embedded or edge ML in the metaverse involves deploying ML models directly on user devices or edge computing infrastructure, minimizing the need to send sensitive data to centralized servers for processing and providing privacy preservation benefits. This approach enables local data processing, reducing communication overhead, latency, and the risk of interception during transmission, thereby enhancing data privacy. By processing data locally, embedded ML enables stronger anonymization techniques and protects user identities and sensitive information by sending only aggregated or anonymized data to central servers. However, resource constraints on edge devices, such as limited computational resources, memory, or power, must be considered when deploying ML models. Optimizing models for efficient execution on resource-constrained devices is crucial to balance privacy preservation and performance. While embedded or edge ML in the metaverse contributes to data privacy by keeping sensitive information on user devices or edge infrastructure, challenges related to resource limitations, model updates, and ensuring trustworthy edge infrastructure must be addressed. Striking a balance between privacy and utility is essential for providing a practical, safe, and privacy-preserving metaverse experience.

\subsection{Taking a Human-Centric AI Approach}
Adopting a human-centric AI development approach for analyzing and creating metaverse applications brings numerous advantages by prioritizing users' needs, values, and well-being. This approach fosters collaboration among AI researchers, developers, designers, ethicists, psychologists, and other stakeholders, promoting a holistic and multidisciplinary development process. By encouraging open dialogue and knowledge sharing, diverse perspectives and potential impacts of AI in the metaverse can be addressed effectively. This collaborative effort ensures that metaverse applications are designed with a deep understanding of human experiences, promoting user-centricity and enabling the creation of meaningful and inclusive virtual environments.

\subsection{Blackbox Nature of DL-Based AI Models} 
The black-box nature of DL models poses significant challenges for interpretability and increases their vulnerability to privacy attacks in metaverse applications, underscoring the importance of incorporating inherently interpretable and transparent models to address these concerns. In the metaverse, where users interact with DL models that process their data, at the same time, interpretable models can provide transparency regarding data usage and processing. This transparency empowers users to comprehend the privacy implications of their data and make informed decisions about sharing sensitive information. Interpretable models enable users to exert greater control over their data and privacy settings within the metaverse. By understanding how the model operates and which features it relies on, users can make informed choices about data sharing and privacy preferences, thus enhancing user autonomy and consent. Furthermore, interpretable models explain their predictions or decisions, allowing users to comprehend the rationale behind recommendations or actions within the metaverse. Such explanations foster user trust, improve the overall user experience, and create opportunities for meaningful interactions with the model. By incorporating inherently interpretable and transparent models into metaverse applications, user privacy can be better protected, as these models promote user control, consent, and understanding of data processing while enabling the detection and mitigation of privacy attacks. Additionally, they enhance accountability, trust, and regulatory compliance, leading to a more privacy-preserving and user-centric metaverse environment.

\section{Conclusions}
\label{sec:conclusions}

This study addresses the crucial issue of privacy protection in the emerging field of AI-XR-enabled metaverses that leverage Artificial Intelligence (AI) and Extended Reality (XR) technologies. The use of personal data in these environments presents significant privacy risks, making it essential to implement effective measures to ensure that AI-XR metaverses are used in a pro-social manner. Our comprehensive review of techniques has shown that while these risks are substantial, they can be mitigated through appropriate regulations and policies and the integration of privacy-enhancing technologies, algorithmic fairness techniques, and interpretable Machine Learning (ML) models. We explore cutting-edge technologies, including Homomorphic Encryption(HE), Differential Privacy (DP), and blockchain, to ensure a safe and secure metaverse environment. The findings of this study have significant implications for policymakers, industry practitioners, and future research, providing valuable insights into the challenges of privacy and security in AI-XR-enabled metaverses. This study emphasizes the need for proactive measures to safeguard user privacy and security, promote the pro-social use of AI-XR metaverses, and build user trust and confidence in these technologies.

\bibliographystyle{IEEEtran}

\end{document}